\documentclass[12pt]{article}
\usepackage{a4wide} 
\newcommand{\Proof}{\noindent {\sc Proof}. }
\newcommand{\qed}{\hfill${\Box}$}

\RequirePackage[latin1]{inputenc}    
\usepackage[english]{babel}
\usepackage{amsmath}
\usepackage{amsfonts}
\usepackage{amssymb}
\usepackage{xspace}
\usepackage{latexsym}
\usepackage{url}
\usepackage{xspace}
\usepackage{fancyvrb}

\newcommand{\vc}[1]{{\bf #1}}

\newtheorem{theorem}{Theorem}

\newtheorem{definition}[theorem]{Definition}
\newtheorem{lemma}[theorem]{Lemma}
\newtheorem{corollary}[theorem]{Corollary}
\newtheorem{proposition}[theorem]{Proposition}
\newtheorem{example}[theorem]{Example}

\newtheorem{remark}[theorem]{Remark}

\newcommand{\Proofitem}[1]{{\ }\\[-\medskipamount]\noindent \textbf{#1}\;}
\newcommand{\Proofitemm}[1]{{\ }\\[-\medskipamount]\noindent $#1\;$}
\newcommand{\Proofitemf}[1]{\noindent $#1\;$}
\newcommand{\Defitem}[1]{{\ }\\[-\medskipamount]\noindent $#1\;$}
\newcommand{\Defitemf}[1]{\noindent $#1\;$}
\newcommand{\eqdef}{=_{\text{def}}}
\newcommand{\concat}{\cdot}

\newcommand{\pc}{\mathit{pc}}

\newcommand{\Error}{\epsilon}

\newcommand{\hbra}{\noindent\hbox to \textwidth{\leaders\hrule height1.8mm depth-1.5mm\hfill}}
\newcommand{\hket}{\noindent\hbox to \textwidth{\leaders\hrule height0.3mm\hfill}}
\newcommand{\ratio}{.3}

\newcommand{\cl}[1]{{\mathcal #1}}

\newcommand{\arrow}{\rightarrow}        
\newcommand{\Z}{{\bf Z}}
\newcommand{\Real}{{\bf R}^{+}}

\newcommand{\Alt}{ \mid\!\!\mid  }
\newcommand{\infer}[2]{\begin{array}{c} #1 \\ \hline #2 \end{array}}
\newcommand{\Arrow}{\Rightarrow}

\newcommand{\union}{\cup}               
               
\newcommand{\Union}{\bigcup}

\newcommand{\comp}{\circ}              
\newcommand{\set}[1]{\{#1\}}

\newcommand{\w}[1]{{\it #1}}

\newcommand{\s}[1]{{\sf #1}}

\newcommand{\act}[1]{\stackrel{#1}{\rightarrow}} 
\newcommand{\eval}{\Downarrow}
\newcommand{\Return}{\ensuremath{\mathtt{return}}\xspace}                
\newcommand{\Stop}{\ensuremath{\mathtt{stop}}\xspace}
\newcommand{\Wait}{\ensuremath{\mathtt{wait}}\xspace}
\newcommand{\Read}{\ensuremath{\mathtt{read}}\xspace}
\newcommand{\Write}{\ensuremath{\mathtt{write}}\xspace}
\newcommand{\Yield}{\ensuremath{\mathtt{yield}}\xspace}
\newcommand{\Next}{\ensuremath{\mathtt{next}}\xspace}
\newcommand{\Load}{\ensuremath{\mathtt{load}}\xspace}
\newcommand{\Call}{\ensuremath{\mathtt{call}}\xspace}
\newcommand{\Tcall}{\ensuremath{\mathtt{tcall}}\xspace}

\newcommand{\Build}{\ensuremath{\mathtt{build}}\xspace}
\newcommand{\Branch}{\ensuremath{\mathtt{branch}}\xspace}

\newcommand{\hatt}[1]{#1^{+}}
\newcommand{\Of}{\mathbin{\w{of}}}
\newcommand{\With}{\mathbin{\w{with}}}
\newcommand{\match}[4]{\mathit{match}\,{#1}\,\mathit{with}\,{#2}
  \,\mathit{then}\,{#3}\,\mathit{else}\,{#4}}
\newcommand{\rmatch}[1]{{\it read}~#1~{\it with}~}

\begin{document}

\title{Resource Control for Synchronous\\ 
Cooperative Threads\thanks{Work partially 
supported by ACI {\em S\'ecurit\'e Informatique} CRISS.}}

\author{
Roberto M. Amadio\\
Universit\'e Paris 7
\thanks{Laboratoire {\em Preuves, Programmes et Syst\`emes}, UMR-CNRS 7126.}
\and 
Silvano {Dal Zilio}\\
CNRS Marseille\thanks{Laboratoire d'Informatique Fondamentale de Marseille, UMR-CNRS 6166}
}

\maketitle

  \begin{abstract}
    We develop new methods to statically bound the resources needed
    for the execution of systems of concurrent, interactive threads.
    Our study is concerned with a \emph{synchronous} model of
    interaction based on cooperative threads whose execution proceeds
    in synchronous rounds called instants. Our contribution is a
    system of compositional static analyses to guarantee that each
    instant terminates and to bound the size of the values computed by
    the system as a function of the size of its parameters at the
    beginning of the instant.\\
    Our method generalises an approach designed for first-order
    functional languages that relies on a combination of standard
    termination techniques for term rewriting systems and an analysis
    of the size of the computed values based on the notion of
    quasi-interpretation.
    We show that these two methods can be combined to obtain an
    explicit polynomial bound on the resources needed for the
    execution of the system during an instant.\\
    As a second contribution, we introduce a virtual machine and a
    related bytecode thus producing a precise description of the
    resources needed for the execution of a system. In this context,
    we present a suitable control flow analysis that allows to
    formulate the static analyses for resource control at byte code
    level.
  \end{abstract}

\section{Introduction}
The problem of bounding the usage made by programs of their resources
has already attracted considerable attention. Automatic extraction of
resource bounds has mainly focused on (first-order) functional
languages starting from Cobham's characterisation~\cite{Cobham65} of
polynomial time functions by bounded recursion on notation. Following
work, see e.g. \cite{BC92,Hofmann02,Jones97,Leivant94}, has developed
various inference techniques that allow for efficient analyses while
capturing a sufficiently large range of practical algorithms.\\
Previous work~\cite{BMM01,Marion00} has shown that polynomial time or
space bounds can be obtained by combining traditional termination
techniques for term rewriting systems with an analysis of the size of
computed values based on the notion of quasi-interpretation.  Thus, in
a nutshell, resource control relies on termination and bounds on data
size.\\
This approach to resource control should be contrasted with
traditional {\em worst case execution time} technology (see,
e.g.,~\cite{RT00}): the bounds are less precise but they apply to a
larger class of algorithms and are {\em functional} in the size of the
input, which seems more appropriate in the context of the applications
we have in mind (see below). In another direction, one may compare the
approach with the one based on linear logic (see, e.g.,~\cite{BM04}):
while in principle the linear logic approach supports higher-order
functions,
it does not offer yet a user-friendly programming language.\\
In~\cite{Amadio02,Amadio04}, we have considered the problem of
automatically inferring quasi-interpretations in the space of
multi-variate max-plus polynomials.  In~\cite{ACDJ04}, we have
presented a virtual machine and a corresponding bytecode for a
first-order functional language and shown how size and termination
annotations can be formulated and verified at the level of the
bytecode. In particular, we can derive from the verification an
explicit polynomial bound on the space required to execute a given
bytecode.\\
In this work, we aim at extending and adapting these results to a
concurrent framework. As a starting point, we choose a basic model of
parallel threads interacting on shared variables.  The kind of
concurrency we consider is a {\em cooperative} one.  This means that
by default a running thread cannot be preempted unless it explicitly
decides to return the control to the scheduler.  In {\em preemptive}
threads, the opposite hypothesis is made: by default a running thread
can be preempted at any point unless it explicitly requires that a
series of actions is atomic. We refer to, e.g.,~\cite{Ous96} for an
extended comparison of the cooperative and preemptive models.  Our
viewpoint is pragmatic: the cooperative model is closer to the
sequential one and many applications are easier to program in the
cooperative model than in the preemptive one. Thus, as a first step,
it makes sense to develop a resource control analysis for the
cooperative model.\\
The second major design choice is to assume that the computation is
regulated by a notion of {\em instant}. An instant lasts as long as a
thread can make some progress in the current instant. In other terms,
an instant ends when the scheduler realizes that all threads are
either stopped, or waiting for the next instant, or waiting for a
value that no thread can produce in the current instant.  Because of
this notion of instant, we regard our model as {\em synchronous}.
Because the model includes a logical notion of time, it is possible
for a thread to react to the absence of an event.\\
The reaction to the absence of an event is typical of synchronous
languages such as \textsc{Esterel}~\cite{BG92}.  Boussinot {\em et
  al.} have proposed a weaker version of this feature where the
reaction to the absence happens in the following instant~\cite{BD92}
and they have implemented it in various programming environments based
on \textsc{C}, \textsc{Java}, and \textsc{Scheme}~\cite{mimosarp}.
Applications suited to this programming style include: event-driven
applications, graphical user interfaces, simulations (e.g. $N$-bodies
problem, cellular automata, ad hoc networks), web services,
multiplayer online games, \dots Boussinot \emph{et al.}  have also
advocated the relevance of this concept for the programming of mobile
code and demonstrated that the possibility for a `synchronous' mobile
agent to react to the absence of an event is an added factor of
flexibility for programs designed for open distributed systems, whose
behaviours are inherently difficult to predict. These applications
rely on data structure such as lists and trees whose size needs to
be controlled.\\
Recently, Boudol~\cite{Boudol04} has proposed a formalisation of this
programming model. Our analysis will essentially focus on a small
fragment of this model without higher-order functions, and where the
creation of fresh memory cells (registers) and the spawning of new
threads is only allowed at the very beginning of an instant.  We
believe that what is left is still expressive and challenging enough
as far as resource control is concerned.  Our analysis goes in three
main steps.  A first step is to guarantee that each instant terminates
(Section~\ref{termination}).  A second step is to bound the size of
the computed values as a function of the size of the parameters at the
beginning of the instant (Section~\ref{size}). A third step, is to
combine the termination and size analyses. 
Here we show how to obtain
polynomial bounds on the {\em space} and {\em time} needed for the execution of the
system during an instant as a function of the size of the parameters
at the beginning of the instant (Section~\ref{size-termination}).  \\
A characteristic of our static analyses is that to a great extent they
make abstraction of the memory and the scheduler. This means that each
thread can be analysed separately, that the complexity of the analyses
grows linearly in the number of threads, and that an incremental
analysis of a dynamically changing system of threads is possible.
Preliminary to these analyses, is a control flow analysis
(Section~\ref{cfa}) that guarantees that each thread performs each
read instruction (in its body code) at most once in an instant. 
This condition is instrumental to resource control.
In particular, it allows to regard behaviours as {\em functions} of their
initial parameters and the registers they may read in the instant.
Taking this functional viewpoint, we are able to adapt the main
techniques developed for proving termination and size bounds in the
first-order functional setting.\\
We point out that our static size analyses are {\em not} intended to
predict the size of the system after arbitrarily many instants.  This
is a harder problem which in general requires an understanding of the
{\em global} behaviour of the system and/or stronger restrictions on
the programs we can write. For the language studied in this paper, we
advocate a combination of our static analyses with a dynamic
controller that at the end of each instant checks the size of the
parameters of the system and may decide to stop some threads taking
too much space.\\
Along the way and in appendix~\ref{examples}, we provide a number of
programming examples illustrating how certain synchronous and/or
concurrent programming paradigms can be represented in our model.
These examples suggest that the constraints imposed by the static
analyses are not too severe and that their verification can be
automated.\\
As a second contribution, we describe a virtual machine and the
related bytecode for our programming model (Section~\ref{vm}).  This
provides a more precise description of the resources needed for the
execution of the systems we consider and opens the way to the
verification of resource bounds at the bytecode level, following the
`typed assembly language' approach adopted in~\cite{ACDJ04} for the
purely functional fragment of the language.  More precisely, we
describe a control flow analysis that allows to recover the conditions
for termination and size bounds at bytecode level and we show that the
control flow analysis is sufficiently liberal to accept the code
generated by a rather standard compilation function.\\
Proofs are available in appendix~\ref{proofs}.


\section{A Model of Synchronous Cooperative Threads}\label{sct}
A {\em system} of synchronous cooperative threads is described by
(1) a list of mutually recursive type and constructor definitions 
and (2) a list of mutually recursive function and behaviour definitions
relying on pattern matching. In this respect, the resulting
programming language is reminiscent of \textsc{Erlang}~\cite{AVWW96},
which is a {\em practical} language to develop concurrent
applications. The set of instructions a behaviour can execute is
rather minimal. Indeed, our language can be regarded as an {\em
  intermediate code} where, for instance, general pattern-matching has
been compiled into a nesting of $\mathit{if\_then\_else}$ constructs
and complex control structures have been compiled into a simple
tail-recursive
form.
\paragraph*{Types}
We denote {\em  type names} with $t, t', \ldots$ and 
{\em constructors} with $\s{c}, \s{c'}, \ldots$ 
We will also denote with $\s{r}, \s{r'}, \ldots$ 
constructors of arity $0$ and of `reference' type (see equation of
kind (2) below) and we will refer to them as {\em registers} (thus
registers are constructors).
The {\em values} $v,v',\ldots$ computed by programs 
are first order terms built out of constructors. 
Types and constructors are declared via recursive equations that may
be of two kinds: 
\[
\begin{array}[c]{c@{\qquad\quad}lcl}
\text{(1)} & t & = & \ldots \mid \s{c} \Of t_1, \ldots, t_n \mid \ldots\\
\text{(2)} & t & = & \w{Ref}(t') \With \ldots \mid \s{r} = v \mid \ldots
\end{array}
\]
In (1) we declare a type $t$ with a constructor $\s{c}$ of functional
type $(t_1,\ldots,t_n) \to t$. In (2) we declare a type $t$ of
registers referencing values of type $t'$ and a register $\s{r}$ with
initial value $v$.  As usual, type definitions can be mutually
recursive (functional and reference types can be intermingled) and it
is assumed that all types and constructors are declared exactly once.
This means that we can associate a unique type with every constructor
and that with respect to this association we can say when a value is
well-typed. For instance, we may define the type $\w{nat}$ of natural
numbers in unary format by the equation $\w{nat} = \s{z} \ \Alt\ \s{s}
\Of \w{nat}$ and the type $\w{llist}$ of linked lists of natural
numbers by the equations $\w{nlist} = \s{nil} \ \Alt\ \s{cons} \Of
(\w{nat}, \w{llist})$ and $\w{llist} = \w{Ref}(\w{nlist}) \With \s{r}
= \s{cons}(\s{z} , \s{r})$. The last definition declares a register
$\s{r}$ of type $\w{llist}$ with initial
value the infinite (cyclic) list containing only $\s{z}$'s.\\
Finally, we have a special {\em behaviour type}, $\w{beh}$.  Elements
of type \w{beh} do not return a value but produce side
effects. We denote with $\beta$ either a regular type or $\w{beh}$.
\paragraph*{Expressions}
We let $x,y,\ldots$ denote {\em variables} ranging over values. 
The {\em size} $|v|$ of a value $v$ is defined by $|\s{c}| = 0$ and
$|\s{c}(v_1, \dots, v_n)| = 1 + |v_1| + \dots + |v_n|$.
In the following, we will use the {\em vectorial notation} 
$\vc{a}$ to denote  either a vector $a_1,\ldots,a_n$ 
or a sequence $a_1 \cdots a_n$ of elements.
We use $\sigma, \sigma', \ldots$ to denote a {\em substitution}
$[\vc{v}/\vc{x}]$, where $\vc{v}$ and $\vc{x}$ have the same length.
A {\em pattern} $p$ is a well-typed term built out of constructors and
variables. In particular, a {\em shallow linear pattern} $p$ is a
pattern $\s{c}(x_1,\ldots,x_n)$, where $\s{c}$ is a constructor of
arity $n$ and the variables $x_1,\ldots,x_n$ are all distinct. 
{\em Expressions}, $e$, and {\em expression bodies}, $\w{eb}$, are defined
as:
\[
\begin{array}[c]{rcl}
  e & ::= & x \ \Alt \
  \s{c}(e_1, \dots, e_k) \ \Alt\
  f(e_1, \dots, e_n)\\
  \w{eb} & ::= & e \ \Alt\ \match{x}{p}{eb}{eb}
\end{array}
\]
\noindent where $f$ is a functional symbol of type $(t_1,\ldots,t_n)
\arrow t$, specified by an equation of the kind $f(x_1, \dots,
x_n)=\w{eb}$, and where $p$ is a shallow linear pattern.\\
A closed expression body $\w{eb}$ evaluates to a value $v$ according
to the following standard rules:
{\small
  \[ 
  \begin{array}{@{}l}

    (\s{e}_{1})~\infer{}{\s{r}\eval \s{r}}

    \qquad

    (\s{e}_{2})~\infer{\vc{e}\eval \vc{v}}
    {\s{c}(\vc{e}) \eval \s{c}(\vc{v})} 

    \qquad
    (\s{e}_3)~\infer{\vc{e} \eval \vc{v},\quad f(\vc{x})=\w{eb},\quad
      [\vc{v}/\vc{x}]\w{eb} \eval v}
    {f(\vc{e})\eval v} \\[2em]

    (\s{e}_4)~\infer{[\vc{v}/\vc{x}]\w{eb}_{1}\eval v}
    {\left(
        \begin{array}{@{}l}
          \w{match}~\s{c}(\vc{v})~\w{with}~\s{c}(\vc{x}) \\
          \w{then}~\w{eb}_{1}~\w{else}~\w{eb}_{2}
        \end{array}\right) \eval v}
    \qquad

    (\s{e}_5)~\infer{\w{eb}_{2}\eval v \quad \s{c}\neq \s{d}}
    {\left(
        \begin{array}{@{}l}
          \w{match}~\s{c}(\vc{v})~\w{with}~\s{d}(\vc{x}) \\
          \w{then}~\w{eb}_{1}~\w{else}~\w{eb}_{2}
        \end{array} 
      \right) \eval v}
  \end{array}
  \]}
Since registers are constructors, rule $(\s{e}_{1})$ is a special 
case of rule $(\s{e}_{2})$; we keep the rule for clarity.

\paragraph*{Behaviours}
Some function symbols may return a thread behaviour $b, b', \ldots$
rather than a value. In contrast to `pure' expressions, a behaviour
does not return a result but produces {\em side-effects} by reading
and writing registers. A behaviour may also affect the scheduling
status of the thread executing it. 
We denote with $b,b',\ldots$ behaviours defined as follows:
\[
\begin{array}{ll}

  b::= &\w{stop} \Alt   f(\vc{e}) \Alt 
  \w{yield}.b \Alt \w{next}.f(\vc{e}) \Alt \varrho := e.b \Alt\\ 
  &\rmatch{\varrho} p_1\Arrow b_1 \mid \cdots \mid p_n \Arrow b_n \mid
  [\_] \Arrow f(\vc{e}) \Alt\\
  &\match{x}{\s{c}(\vc{x})}{b_{1}}{b_{2}} 
\end{array}
\]
\noindent where: (i) $f$ is a functional symbol of type $t_1,\ldots,t_n
\arrow \w{beh}$, defined by an equation $f(\vc{x}) = b$, (ii) 
$\varrho, \varrho', \ldots$ range over variables and registers,  and
(iii) $p_1,\ldots,p_n$ are either shallow linear patterns or
variables.  We also denote with $[\_]$ a special symbol that will be used 
in the default case of \w{read} expressions (see the paragraph 
\textbf{Scheduler} below). Note that if the pattern $p_i$ is a variable then 
the following branches including the default one can never be
executed. 

The {\em effect} of the various instructions is informally described
as follows: $\w{stop}$, terminates the executing thread for ever;
$\w{yield}.{b}$, halts the execution and hands over the control to the
scheduler --- the control should return to the thread later in the
same instant and execution resumes with $b$; $f(\vc{e})$ and
$\w{next}.{f(\vc{e})}$ switch to another behaviour immediately or at
the beginning of the following instant; $\s{r}:= e.b$, evaluates the
expression $e$, assigns its value to $\s{r}$ and proceeds with the
evaluation of $b$; $\w{read}\, \s{r}\, \w{with} \, p_1 \Arrow b_1 \mid
\dots \mid p_n \Arrow b_n \mid [\_] \Arrow b$, waits until the value
of $\s{r}$ matches one of the patterns $p_1, \dots, p_n$ (there could
be no delay) and yields the control otherwise; if at the end of the
instant the thread is always stuck waiting for a matching value
then it starts the behaviour $b$ in the following instant;
$\match{v}{p}{b_{1}}{b_{2}}$ filters the value $v$ according to the
pattern $p$, it never blocks the execution. Note that if $p$ is a
pattern and $v$ is a value there is at most one matching substitution
$\sigma$ such that $v = \sigma p$.

Behaviour reduction is described by the 9 rules below. A reduction
$(b,s) \smash[t]{\stackrel{X}{\rightarrow}} (b',s')$ means that the
behaviour $b$ with store $s$ runs an atomic sequence of actions till
$b'$, producing a store $s'$, and returning the control to the
scheduler with status $X$. A status is a value in $\set{N,R,S,W}$ that
represents one of the four possible state of a thread --- $N$ stands
for next (the thread will resume at the beginning of the next
instant), $R$ for run, $S$ for stopped, and $W$ for wait (the thread
is blocked on a $\w{read}$ statement).
{\small
  \[
  \begin{array}{@{}l}

    (\s{b}_1)~\infer{~}
    {(\w{stop}, s) \act{S} (\w{stop}, s)} 

    ~

    (\s{b}_2)~\infer{}{(\w{yield}.{b}, s) \act{R} (b, s)} 
    
    ~
    
    (\s{b}_3)~\infer{}{ (\w{next}.{ f(\vc{e}) }, s) \act{N} (f(\vc{e}),
      s)}\\[2em]

    (\s{b}_4)~\infer{([\vc{v}/\vc{x}]b_{1},s) \act{X} (b',s')}
    {\left(
        \begin{array}{@{}c}
          \w{match}~\s{c}(\vc{v})\\
          \w{with}~\s{c}(\vc{x}) \\
          \w{then}~b_{1}~\w{else}~b_{2}
        \end{array},~s \right) \act{X} (b',s')}
    \
    (\s{b}_5)~\infer{(b_{2},s)\act{X} (b',s'),\quad \s{c}\neq \s{d}}
    {\left(
        \begin{array}{@{}c}
          \w{match}~\s{c}(\vc{v})\\
          \w{with}~\s{d}(\vc{x}) \\
          \w{then}~b_{1}~\w{else}~b_{2}
        \end{array} ,~ s\right) \act{X} (b',s')} \\[2em]
    
    (\s{b}_6)~\infer{\mbox{no pattern matches }\w{s}(\s{r})}
    {(\w{read}~\s{r}\ldots,s )\act{W}
      (\w{read}~\s{r}\ldots,s)} 
    
    ~~
    
    (\s{b}_7)~\infer{s(\s{r})= \sigma p,\quad (\sigma b,s)\act{X}(b',s')}
    {(\w{read}~\s{r}~\w{with} \dots \mid p \Arrow b \mid \dots,s)\act{X}(b',s')} \\[2em]

    (\s{b}_8)~
    \infer{\vc{e}\eval \vc{v},\quad f(\vc{x})=b, \quad
      ([\vc{v}/\vc{x}]b,s)\act{X} (b',s')}
    {(f(\vc{e}),s) \act{X} (b',s')} 
    
    \qquad
    
    (\s{b}_9)~\infer{e\eval v,\quad(b,s[v/\s{r}]) \act{X} (b',s')}
    {(\s{r}:=e.b,s) \act{X} (b',s')}
    
  \end{array}
  \]}
We denote with $\w{be}$ either an expression body or a
behaviour. All expressions and behaviours are supposed to be
{\em well-typed}.  As usual, all formal parameters are supposed to be
distinct. In the $\mathit{match}\,{x}\,\mathit{with}\,{\s{c}(\vc{y})}
\,\mathit{then}\,{\w{be}_{1}}$ $\mathit{else}\,{\w{be}_{2}}$
instruction, $\w{be}_{1}$ may depend on $\vc{y}$ but not on $x$ while
$\w{be}_{2}$ may depend on $x$ but not on $\vc{y}$.

\paragraph*{Systems}
We suppose that the execution environment consists of $n$ threads and
we associate with every thread a distinct identity that is an index in
$\Z_n=\set{0,1,\ldots,n-1}$. We let $B,B',\ldots$ denote 
\emph{systems} of synchronous threads, that is finite mappings from
thread indexes to pairs (behaviour, status).  Each register has a type
and a default value --- its value at the beginning of an instant ---
and we use $s,s',\dots$ to denote a \emph{store}, an association
between registers and their values. We suppose that at the beginning
of each instant the store is $s_o$, such that each register is
assigned its default value. If $B$ is a system and $i\in\Z_n$ is a
valid thread index then we denote with $B_1(i)$ the behaviour executed
by the thread $i$ and with $B_2(i)$ its current status. Initially, all
threads have status $R$, the current thread index is $0$, and $B_1(i)$
is a behaviour expression of the shape $f(\vc{v})$ for all $i \in
\Z_n$. System reduction is described by a relation $(B,s,i) \arrow
(B',s',i')$: the system $B$ with store $s$ and current thread (index)
$i$ runs an atomic sequence of actions and becomes
$(B',s',i')$.
{\small
\[
\begin{array}{c}

   (\s{s}_1)~\infer{(B_1(i),s) \act{X} (b',s'),\quad 
     B_2(i)=R,\quad
     B'= B[(b',X)/i],\quad \cl{N}(B',s',i) = k}
   {(B,s,i) \arrow (B'[(B'_{1}(k),R)/k],s',k)}\\[1.5em]

   (\s{s}_2)~\infer{\begin{array}{c}
       (B_1(i),s) \act{X} (b',s'),\quad B_2(i)=R,\quad
       B'=B[(b',X)/i],\quad
       \cl{N}(B',s',i) \uparrow,\\
       B''=\cl{U}(B',s'),\quad
       \cl{N}(B'',s_o,0) = k
     \end{array}}
   {(B,s,i) \arrow (B'',s_o,k)}

\end{array}
\]}

\paragraph*{Scheduler}
The scheduler is determined by the functions $\cl{N}$ and $\cl{U}$.
To ensure progress of the scheduling, we assume that if $\cl{N}$
returns an index then it must be possible to run the corresponding
thread in the current instant and that if $\cl{N}$ is undefined
(denoted $\cl{N}(\dots) \uparrow$) then no thread can be run in the
current instant.
{\small
\[
\begin{array}{ll}
     \mbox{If }\cl{N}(B,s,i) = k \mbox{ then}       
     &B_2(k)=R \mbox{ or } (~B_2(k)=W \mbox{ and}\\
     &\quad B_1(k)=\w{read}\,\s{r}\,\w{with}\, \dots \mid p\Arrow b \mid
     \dots \mbox{ and some pattern}\\
     &\quad \mbox{matches }s(\s{r}) \mbox{ {i.e.}, } 
     \exists \sigma \,\sigma p =s(\s{r})~) \\

     \mbox{If }\cl{N}(B,s,i) \uparrow \mbox{ then} 
     &\forall k\in \Z_n,~B_2(k)\in \set{N,S} \mbox{ or }
     (~B_2(k)=W, \\
     &\quad B_1(k)=\w{read}\,\s{r}\,\w{with}\, \dots, \mbox{
       and no pattern matches
     }s(\s{r})~)
   \end{array}
\]}

When no more thread can run, the instant ends and the function
$\cl{U}$ performs the following status transitions: $N\arrow R$,
$W\arrow R$. We assume here that every thread in
status $W$ takes the $[\_] \Arrow \dots$ branch at the beginning of
the next instant. Note that the function $\cl{N}$ is undefined on the
updated system if and only if all threads are stopped.
{\small
\[
   \begin{array}{c}
     \cl{U}(B,s)(i) =        \left\{\begin{array}{ll}
         (b,S)  &\mbox{if }B(i)=(b,S) \\
         (b,R)  &\mbox{if }B(i)=(b,N) \\
         (f(\vc{e}),R) &\mbox{if
         }B(i)=(\w{read}\,\s{r} \,\w{with}\, \dots \mid [\_]
         \Arrow f(\vc{e}),W)
       \end{array}\right.
   \end{array}
\]}

\begin{example}[channels and signals]\label{ex-ch-sig}
  The \w{read} instruction allows to read a register subject to
  certain filter conditions. This is a powerful mechanism which
  recalls, e.g., Linda communication~\cite{CG89}, and that allows to
  encode various forms of channel and signal
  communication.\\
  \Defitem{(1)} We want to represent a {\em one place channel} $\s{c}$
  carrying values of type $t$. We introduce a new type $\w{ch}(t) =
  \s{empty} \mid \s{full}~\w{of}~t$ and a register $\s{c}$ of type
  $\w{Ref}(\w{ch}(t))$ with default value $\s{empty}$. A thread should
  send a message on $\s{c}$ only if $\s{c}$ is empty and it should
  receive a message only if $\s{c}$ is {\em not} empty (a received
  message is discarded). These operations can be modelled using the
  following two
  derived operators:\\
  \[{\small
  \begin{array}{ll}
    \w{send}(\s{c},e).b &\eqdef\ \w{read}~\s{c}~\w{with}\, 
    \s{empty} \Arrow \s{c}:=\s{full}(e).b \\
    
    \w{receive}(\s{c},x).b &\eqdef\ 
    \w{read}~\s{c}~\w{with}~\s{full}(x) \Arrow \s{c}:=\s{empty}.b
  \end{array}}
  \]
  \Defitemf{(2)} We want to represent a {\em fifo channel} \s{c}
  carrying values of type $t$ such that a thread can always emit a
  value on \s{c} but may receive only if there is at least one message
  in the channel. We introduce a new type $\w{fch}(t) = \s{nil} \mid
  \s{cons}~\w{of}~t, \w{fch}(t)$ and a register $\s{c}$ of type
  $\w{Ref}(\w{fch}(t))$ with default value $\s{nil}$. Hence a fifo channel is
  modelled by a register holding a list of values. We consider two
  read operations --- \w{freceive} to fetch the first message on the
  channel and \w{freceiveall} to fetch the whole queue of messages ---
  and we use the auxiliary function \emph{insert} to queue messages at
  the end of the list:
  \[{\small
  \begin{array}{l@{\ }c@{~~}l}

    \w{fsend}(\s{c},e).b &\eqdef& 
    \w{read}~\s{c}~\w{with}~l \Arrow \s{c}:=\w{insert}(e,l).b \\
    
    \w{freceive}(\s{c},x).b  &\eqdef& 
    \w{read}~\s{c}~\w{with}~\s{cons}(x,l) \Arrow \s{c}:=l.b  \\
    
    \w{freceiveall}(\s{c},x).b &\eqdef& 
    \w{read}~\s{c}~\w{with}~\s{cons}(y,l)\Arrow
    \s{c}:=\s{nil}.[\s{cons}(y,l)/x]b \\[0.4em]

    \w{insert}(x,l) &=& 
    \w{match}~l~\w{with}~\s{cons}(y,l')~
    \w{then} ~\s{cons}(y, \w{insert}(x, l'))\\
    && \w{else} ~ \s{cons}(x, \s{nil}) 
  \end{array}}
  \]
\Defitemf{(3)} We want to represent a signal \s{s} with the typical
  associated primitives: emitting a signal and blocking until a signal
  is present. We define a type $\w{sig}= \s{abst} \mid \s{prst}$ and a
  register \s{s} of type $\w{Ref}(\w{sig})$ with default value $\s{abst}$,
  meaning that a signal is originally absent:
  \[{\small
  \begin{array}{l@{\qquad\qquad}l}
    \w{emit}(\s{s}).b \ \eqdef\ \s{s}:=\s{prst}.b 
    &\w{wait}(\s{s}).b \ \eqdef\ 
    \s{read}~\s{s}~\s{with}~ \s{prst} \Arrow b
  \end{array}}
  \]
\end{example}

\begin{example}[cooperative fragment]\label{cooperative}
  The {\em cooperative} fragment of the model with no synchrony is
  obtained by removing the \w{next} instruction and assuming that for
  all \w{read} instructions the branch $[\_] \Arrow f(\vc{e})$ is such
  that $f(\dots) = \w{stop}$. Then all the interesting computation
  happens in the first instant; threads still running in the second
  instant can only stop. By using the representation of fifo channels
  presented in Example~\ref{ex-ch-sig}(2) above, the cooperative
  fragment is already powerful enough to simulate, e.g., Kahn
  networks~\cite{Kahn74}.
\end{example}

Next, to make possible a compositional and functional analysis for
resource control, we propose to restrict the admissible behaviours and
we define a simple preliminary control flow analysis that guarantees
that this restriction is met. We then rely on this analysis to define
a symbolic representation of the states reachable by a behaviour.
Finally, we extract from this symbolic control points suitable order
constraints which are instrumental to our analyses for termination and
value size limitation within an instant.

\subsection{Read Once Condition}\label{cfa}
We require and statically check on the \emph{call graph} of the
program (see below) that threads can perform any given read
instruction at most once in an instant.

\begin{enumerate}
  
\item We assign to every read instruction in a system a distinct fresh
  label, $y$, and we collect all these labels in an ordered sequence,
  $y_1,\ldots, y_m$. In the following, we will sometimes use the
  notation $\w{read}_{\langle y\rangle}\, \varrho \ \w{with}\, \dots$ in the
  code of a behaviour to make visible the label of a \w{read}
  instruction.

\item With every function symbol $f$ defined by an equation
  $f(\vc{x})=b$ we associate the set $L(f)$ of labels of read
  instructions occurring in $b$.
  
\item We define a directed {\em call} graph $G=(N,E)$ as follows: $N$
  is the set of function symbols in the program defined by an equation
  $f(\vc{x})=b$ and $(f,g)\in E$ if $g\in \w{Call}(b)$ where
  $\w{Call}(b)$ is the collection of function symbols in $N$ that may
  be called in the current instant and which is formally defined as
  follows:
  \[{\small
  \begin{array}{@{}c}
    \w{Call}(\w{stop})= \w{Call}(\w{next}.g(\vc{e})) = \emptyset 
    \quad
    \w{Call}(f(\vc{e})) = \{f\}\\

    \w{Call}(\w{yield}.b) = \w{Call}(\varrho:=e.b) = \w{Call}(b)\\

    \w{Call}(\match{x}{p}{b_{1}}{b_{2}}) = \w{Call}(b_{1})\union
    \w{Call}(b_{2})\\
 
    \w{Call}(\rmatch{\varrho} p_1\Arrow b_1 \mid \cdots \mid p_n \Arrow b_n \mid [\_] \Arrow b)
    = \Union_{i=1,\ldots,n} \w{Call}(b_{i})
  \end{array}}
  \]
  We write $f E^* g$ if the node $g$ is reachable from the node $f$ in
  the graph $G$. We denote with $R(f)$ the set of labels $\Union
  \set{L(g) \mid f E^* g}$ and with $\vc{y}_f$ the ordered sequence of
  labels in $R(f)$.
\end{enumerate}

The definition of \w{Call} is such that for every sequence of calls in
the execution of a thread within an instant we can find a
corresponding path in the call graph.

\begin{definition}[read once condition]
  A system satisfies the read once condition if in the call graph
  there are no loops that go through a node $f$ such that $L(f)\neq
  \emptyset$.
\end{definition}

\begin{example}[alarm]\label{alarm-function}
  We consider the representation of signals as in
  Example~\ref{ex-ch-sig}(3). We assume two signals $\s{sig}$ and
  $\s{ring}$. The behaviour $\w{alarm}(n, m)$ will emit a signal on
  $\s{ring}$ if it detects that no signal is emitted on $\s{sig}$ for
  $m$ consecutive instants.  The alarm delay is reset to $n$ if the
  signal $\s{sig}$ is present.
  \[{\small
  \begin{array}{ll}
    \w{alarm}(x,y) = & \w{match}~y~\w{with}~ \s{s}(y') \\
    &\w{then}~\w{read}_{\langle u\rangle}\, \s{sig}\, \w{with}\,
    \s{prst} \Arrow \w{next}.{\w{alarm}(x, x) } ~\mid~
    [\_] \Arrow \w{alarm}(x, y')\\
    &\w{else}~\s{ring}:=\s{prst}.\w{stop}
  \end{array}}
  \]
  Hence $u$ is the label associated with the \w{read} instruction and
  $L(\w{alarm})=\set{u}$. Since the call graph has just one node,
  \w{alarm}, and no edges, the read once condition is satisfied.
\end{example}

To summarise, the read once condition is a checkable syntactic
condition that safely approximates the semantic property we are aiming
at.

\begin{proposition}\label{read-once-prop}
If a system satisfies the read once condition
then in every instant every thread runs every read instruction 
at most once (but the same read instruction can be run by several
threads).
\end{proposition}

The following simple example shows that {\em without} the
read once restriction, a thread can use a register as an accumulator
and produce an exponential growth of the size of the data within an
instant.

\begin{example}[exponentiation]\label{example:exponent}
  We recall that $\w{nat}= \s{z} \mid \s{s}\,\w{of}\,\w{nat}$ is the
  type of tally natural numbers. The function \emph{dble} defined
  below doubles the value of its parameter so that
  $|\w{dble}(n)|=2|n|$.  We assume $\s{r}$ is a register of type
  $\w{nat}$ with initial value $\s{s}(\s{z})$. Now consider the
  following recursive behaviour:
  \[{\small
  \begin{array}{ll}
    \w{dble}(n) = &\w{match}~n~\w{with}~\s{s}
    (n')~\w{then}~\s{s}(\s{s}(\w{dble}(n')))~\w{else}~\s{z}~\\[0.4em]

    \w{exp}(n) = &\w{match}~n~\w{with}~\s{s}(n') \\
    &\w{then}~\w{read}\, \s{r} \,\w{with}\, 
    m \Arrow \s{r}:=\w{dble}(m).\w{exp}(n') \\
    &\w{else}~ \w{stop}
  \end{array}}
  \]
  The function $\w{exp}$ does {\em not} satisfy the read once
  condition since the call graph has a loop on the \w{exp} node.  The
  evaluation of $\w{exp}(n)$ involves $|n|$ reads to the register
  $\s{r}$ and, after each read operation, the size of the value stored
  in $\s{r}$ doubles. Hence, at end of the instant, the register
  contains a value of size $2^{|n|}$.
\end{example}

The read once condition does not appear to be a severe limitation on
the expressiveness of a synchronous programming language.
Intuitively, in most synchronous algorithms every thread reads some
bounded number of variables before performing some action. Note that
while the number of variables is bounded by a constant, the amount of
information that can be read in each variable is not. Thus, for
instance, a `server' thread can just read one variable in which is
stored the list of requests produced so far and then it can go on
scanning the list and replying to all the requests within the same
instant.

\subsection{Control Points}\label{sec:control-points}
From a technical point of view, an important consequence
of the {\em read once} condition is that a behaviour can be described
as a {\em function} of its parameters and the registers it may read
during an instant. This fact is used to associate with a system
satisfying the
read once condition a {\em finite} number of control points.\\
A {\em control point} is a triple $(f(\vc{p}),\w{be},i)$ where,
intuitively, $f$ is the currently called function, $\vc{p}$ represents
the patterns crossed so far in the function definition plus possibly
the labels of the read instructions that still have to be executed,
$\w{be}$ is the continuation, and $i$ is an integer flag in $\{0, 1,
2\}$ that will be used to associate with the control point various
kinds of conditions.\\
If the function $f$ returns a value and is defined by the equation
$f(\vc{x})= \w{eb}$, then we associate with $f$ the set
$\cl{C}(f,\vc{x},\w{eb})$ defined as follows:\\
\[{\small
\begin{array}{ll}
\multicolumn{2}{l}{\cl{C}(f,\vc{p},\w{eb}) = \w{case}~\w{eb}~\w{of}}\\
\ e &: \set{(f(\vc{p}),\w{eb},0)} \\

\left(
\begin{array}{@{}l@{}}
\w{match}~x~\w{with}~\s{c}(\vc{y}) \\
\w{then}~\w{eb}_{1}~\w{else}~\w{eb}_{2}
\end{array}
\right)
&:
\set{(f(\vc{p}), \w{eb}, 2)} \union
\cl{C}(f,[\s{c}(\vc{y})/x]\vc{p},\w{eb}_{1})
\union
\cl{C}(f,\vc{p},\w{eb}_{2})
\end{array}}
\]
On the other hand, suppose the function $f$ is a behaviour defined by
the equation $f(\vc{x})= b$. Then we generate a fresh function symbol
$\hatt{f}$ whose arity is that of $f$ plus the size of $R(f)$, thus
regarding the labels $\vc{y}_f$ (the ordered sequence of labels in
$R(f)$) as part of the formal parameters of $\hatt{f}$.  The set of
control points associated with $\hatt{f}$ is the set $\cl{C}(\hatt{f},
(\vc{x} \cdot \vc{y}_f), b)$ defined as follows:\\
{\small
\begin{tabular}{lll}

\multicolumn{3}{l}{$\cl{C}(\hatt{f}, \vc{p}, b) =  \w{case}~ b~\w{of }$}\\

$(\cl{C}_{1})$  &$\w{stop}$  &$: \set{(\hatt{f}(\vc{p}), b, 2)}$ \\

$(\cl{C}_{2})$  &$g(\vc{e})$ &$: \set{(\hatt{f}(\vc{p}), b, 0)}$ \\ 
  
$(\cl{C}_{3})$  &$\w{yield}.{b'}$ &$: 
                \set{(\hatt{f}(\vc{p}), b, 2)} \union 
                \cl{C}(\hatt{f}, \vc{p} , b')$ \\

$(\cl{C}_{4})$  &$\w{next}.{g(\vc{e})}$
&$: 
\set{(\hatt{f}(\vc{p}), b, 2),
    (\hatt{f}(\vc{p}), g(\vc{e}), 2)}$ \\

$(\cl{C}_{5})$  &$\varrho := e . b'$
&$:
\set{(\hatt{f}(\vc{p}), b, 2), 
(\hatt{f}(\vc{p}), e, 1)} \union
\cl{C}(\hatt{f}\!\!, \vc{p} , b')$ \\
  
\end{tabular}}

{\small
\begin{tabular}{lll}

$(\cl{C}_{6})$  &$
\left(
\begin{array}{@{}l@{}}
\w{match}~x~\w{with}~\s{c}(\vc{y}) \\
\w{then}~\w{b}_{1}~\w{else}~\w{b}_{2}
\end{array}
\right)$
&$:
\begin{array}{@{}l@{}}
\set{(\hatt{f}(\vc{p}), b,2)}\union
\cl{C}(\hatt{f}, ([\s{c}(\vc{y})/x]\vc{p}), \w{b}_{1})\\
~\union~
\cl{C}(\hatt{f},\vc{p},\w{b}_{2})
\end{array}$ \\

$(\cl{C}_{7})$  &$
\left(\begin{array}{@{}l@{}}
\w{read}_{\langle y\rangle}\, \varrho \,\w{with}\, p_1 \Arrow b_1 \mid \dots \mid \\
 p_n \Arrow b_n \mid [\_] \Arrow g(\vc{e}) 
\end{array}\right)$
&$:
\left.\begin{array}{@{}l}
\set{(\hatt{f}(\vc{p}), b, 2 ),
   (\hatt{f}(\vc{p}), g(\vc{e}), 2)} \\
 ~\union~ \cl{C}(\hatt{f}, ([p_1/y]\vc{p}) , b_1) \ \union
 \dots\\
 ~\union~
 \cl{C}(\hatt{f}\!\!, ([p_n/y]\vc{p}) , b_n)
\end{array}\right.$
\end{tabular}}

By inspecting the definitions, we can check that a control point
$(f(\vc{p}), \w{be}, i)$ has the property that
$\w{Var}(\w{be})\subseteq \w{Var}(\vc{p})$.
\begin{definition}
  An instance of a control point $(f(\vc{p}),\w{be},i)$ is an
  expression body or a behaviour $\w{be}'=\sigma (\w{be})$, where
  $\sigma$ is a substitution mapping the free variables in $\w{be}$ to
  values.
\end{definition}

The property of being an instance of a control point is preserved by
expression body evaluation, behaviour reduction and system reduction.
Thus the control points associated with a system do provide a
representation of all reachable configurations. Indeed, in
Appendix~\ref{proofs} we show that it is possible to define the
evaluation and the reduction on pairs of control points and
substitutions.
\begin{proposition}\label{prop-cp}
  Suppose $(B,s,i) \arrow (B',s',i')$ and that for all thread indexes
  $j\in\Z_n$, $B_1(j)$ is an instance of a control point. Then for all
  $j\in\Z_n$, we have that $B'_1(j)$ is an instance of a control
  point.
\end{proposition}
In order to prove the termination of the instant and to obtain a bound
on the size of computed value, we
associate order constraints with control points:
\[{\small
\begin{array}{@{\qquad}l@{\qquad}|@{\qquad}l@{\qquad}}
  
  \mbox{Control point}            & \mbox{Associated constraint}
  \\\hline

  (f(\vc{p}),e,0)                  & f(\vc{p})\succ_0 e \\
  (\hatt{f}(\vc{p}),g(\vc{e}),0)   & \hatt{f}(\vc{p}) \succ_0 \hatt{g}(\vc{e},\vc{y}_g) \\
  (\hatt{f}(\vc{p}),e,1)           & \hatt{f}(\vc{p})\succ_1 e \\
  (\hatt{f}(\vc{p}),\w{be},2)      & \mbox{\emph{no constraints} }
\end{array}}
\]

A program will be deemed correct if the set of constraints obtained
from all the function definitions can be satisfied in suitable
structures.  We say that a constraint $e\succ_i e'$ has index $i$. We
rely on the constraints of index $0$ to enforce termination of the
instant and on those of index $0$ or $1$ to enforce a bound on the
size of the computed values.  Note that the constraints are on pure
first order terms, a property that allows us to reuse techniques
developed in the standard term rewriting framework (cf.
Section~\ref{res-contr}).
\begin{example}\label{constr-alarm-ex}
  With reference to Example~\ref{alarm-function}, we obtain the
  following control points:
  \[{\small
  \begin{array}{l@{\ \:}l}

    (\hatt{\w{alarm}}(x, y, u), \w{match}\, \dots, 2) 
    & (\hatt{\w{alarm}}(x, y, u), \s{ring}:=\s{prst}.\w{stop}, 2)\\

    (\hatt{\w{alarm}}(x, y, u), \s{prst}, 1)
    & (\hatt{\w{alarm}}(x, \s{z}, u), \w{stop}, 2)\\

    (\hatt{\w{alarm}}(x, \s{s}(y'), u) , \w{read}\, \dots, 2)
    & (\hatt{\w{alarm}}(x, \s{s}(y'), u) , \w{alarm}(x, y'), 2)\\

    (\hatt{\w{alarm}}(x, \s{s}(y'), \s{prst}), \w{next}. {\w{alarm}(x, x)}, 2)
    & (\hatt{\w{alarm}}(x, \s{s}(y'), \s{prst}), \w{alarm}(x, x), 2)
    \\

  \end{array}}
  \]
  The triple $(\hatt{\w{alarm}}(x, y, u), \s{prst}, 1)$ is the only
  control point with a flag different from $2$. It corresponds to the
  constraint ${\hatt{\w{alarm}}}(x, y, u) \succ_1 \s{prst}$, where $u$
  is the label associated with the only \w{read} instruction in the
  body of \w{alarm}.  We note that no constraints of index $0$ are
  generated and so, in this simple case, the control flow analysis can
  already establish the termination of the thread and all is left to
  do is to check that the size of the data is under control, which is
  also easily verified.
\end{example}

In Example~\ref{cooperative}, we have discussed a possible
representation of Kahn networks in the cooperative fragment of our
model.  In general Kahn networks there is no bound on the number of
messages that can be written in a fifo channel nor on the size of the
messages. Much effort has been put into the {\em static scheduling} of
Kahn networks (see, e.g.,~\cite{LeeMer87,Caspi92,CaspiPouzet96}). This
analysis can be regarded as a form of resource control since it
guarantees that the number of messages in fifo channels is bounded
(but says nothing about their size). The static scheduling of Kahn
network is also motivated by performance issues, since it eliminates
the need to schedule threads at run time. Let us look in some detail
at the programming language \textsc{Lustre}, that can be regarded as a
language for programming Kahn networks that can be executed {\em
  synchronously}.

\begin{example}[read once vs. \textsc{Lustre}]
  A \textsc{Lustre} network is composed of four types of nodes: the
  combinatorial node, the delay node, the when node, and the merge
  node.  Each node may have several input streams and one output
  stream. The functional behaviour of each type of node is defined by
  a set of recursive definitions. For instance, the node $\w{When}$
  has one boolean input stream $b$ --- with values of type $\w{bool} =
  \s{false} \mid \s{true}$ --- and one input stream $s$ of values. A
  \emph{When} node is used to output values from $s$ whenever $b$ is
  true. This behaviour may be described by the following recursive
  definitions: $\w{When}(\s{false} \cdot b, x \cdot s) = \w{When}(b,
  s)$, $\w{When}(\s{true} \cdot b, x \cdot s) = x \cdot \w{When}(b,
  s)$, and $\w{When}(b, s) = \epsilon$ otherwise. Here is a possible
  representation of the $\w{When}$ node in our model, where the input
  streams correspond to one place channels $\s{b},\s{c}$ (cf.
  Example~\ref{ex-ch-sig}(1)), the output stream to a one place
  channel $\s{c'}$ and at most one element in each input stream is
  processed per instant.
  \[{\small
  \begin{array}{l}
    \w{When}()\ =\  \w{read}_{\langle u\rangle}\, \s{b}~\w{with}  \\
    \quad \begin{array}[t]{c@{\ }l}
      & \s{full}(\s{true}) \Arrow 
      \w{read}_{\langle v\rangle}\, \s{c} \,\w{with}\,
      \s{full}(x) \Arrow \s{c'}:=x.\w{next}.\w{When}() \mid [ \_ ] \Arrow \w{When}()\\
      \mid & \s{full}(\s{false}) \Arrow \w{next}.\w{When}()\\
      \mid & [ \_ ] \Arrow \w{When}()\\
    \end{array}
  \end{array}}
  \]
  While the function $\w{When}$ has no formal parameters, we consider
  the function $\hatt{\w{When}}$ with two parameters $u$ and $v$ in
  our size and termination analyses.
\end{example}
\section{Resource Control}\label{res-contr}
Our analysis goes in three main steps: first, we guarantee that each
instant terminates (Section~\ref{termination}), second we bound the
size of the computed values as a function of the size of the
parameters at the beginning of the instant (Section~\ref{size}), and
third we combine the termination and size analyses to obtain
polynomial bounds on space and time (Section~\ref{size-termination}).\\
%
As we progress in our analysis, we refine the techniques we
employ. Termination is reduced to the general problem of finding a
suitable well-founded order over first-order terms. Bounding the size
of the computed values is reduced to the problem of synthesizing a
{\em quasi-interpretation}.  Finally, the problem of obtaining
polynomial bounds is attacked by combining {\em recursive path
ordering} termination arguments with quasi-interpretations. We selected
these techniques because they are well established and they can handle
a significant spectrum of the programs we are interested in.
It is to be expected that other characterisations of complexity classes 
available in the literature may lead to similar results.

\subsection{Termination of the Instant}\label{termination}
We recall that a {\em reduction order} $>$ over first-order terms is a
well-founded order that is closed under context and substitution:
$t>s$ implies $C[t]>C[s]$ and $\sigma t > \sigma s$, where $C$ is any
one hole context and $\sigma$ is any substitution (see, {\em
  e.g},~\cite{BaaderNipkow98}).

\begin{definition}[termination condition]
  We say that a system satisfies the termination condition if there is
  a reduction order $>$ such that all constraints of index $0$
  associated with the system hold in the reduction order.
\end{definition}

In this section, we assume that the system satisfies the termination
condition.  As expected this entails that the evaluation of closed
expressions succeeds.

\begin{proposition}\label{exp-eval}
  Let $e$ be a closed expression. Then there is a value $v$ such that
  $e\eval v$ and $e \geq v$ with respect to the reduction order.
\end{proposition}
 
Moreover, the following proposition states that a behaviour will always
return the control to the scheduler.

\begin{proposition}[progress]\label{beh-red}
  Let $b$ be an instance of a control point.  Then for all stores $s$,
  there exist $X, b'$ and $s'$ such that
  $\smash[t]{(b,s)\act{X}(b',s')}$.
\end{proposition}

Finally, we can guarantee that at each instant the system will reach a
configuration in which the scheduler detects the end of the instant
and proceeds to the reinitialisation of the store and the status (as specified
by rule $(\s{s}_2)$).

\begin{theorem}[termination of the instant]\label{sys-ter}
  All sequences of system reductions involving only rule $(\s{s}_1)$
  are finite.
\end{theorem}

Proposition \ref{beh-red} and Theorem \ref{sys-ter} are proven by
exhibiting a suitable well-founded measure which is based both on the
reduction order and the fact that the number of reads a thread may
perform in an instant is finite.

\begin{example}[monitor max value]\label{max-value-ex}
  We consider a recursive behaviour monitoring the register $\s{i}$
  (acting as a fifo channel) and parameterised on a number $x$
  representing the largest value read so far. At each instant, the
  behaviour reads the list $l$ of natural numbers received on \s{i}
  and assigns to \s{o} the greatest number in $x$ and $l$.

  \[{\small
  \begin{array}{lcl}
    f(x)      &=&\w{yield}.\w{read}_{\langle i\rangle}\, 
    \s{i} \,\w{with}\, l \Arrow f_{1}(\w{maxl}(l, x)) \\
    f_{1}(x)  &=&\s{o}:=x.\w{next}.f(x) \\
    \w{max}(x,y) &=&\w{match}\, x \,\w{with}\, \s{s}(x')  \\
    &&\w{then}\, 
    \w{match}\, y \,\w{with}\, \s{s}(y')\,
    \w{then}\, \s{s}(\w{max}(x',y')) \,\w{else}\, \s{s}(x')\\
    &&\w{else}~y \\

    \w{maxl}(l,x) &=&\w{match}\, l \,\w{with}\, \s{cons}(y,l')\,
    \w{then}\, \w{maxl}(l',\w{max}(x,y)) \,\w{else}\, x

  \end{array}}
  \]
  It is easy to prove the termination of the thread by recursive path
  ordering, where the function symbols are ordered as $\hatt{f} >
  \hatt{f}_{1} > \w{maxl} > \w{max}$, the arguments of $\w{maxl}$ are
  compared lexicographically from left to right, and the constructor
  symbols are incomparable and smaller than any function symbol.
\end{example}

\subsection{Quasi-interpretations}\label{size}
Our next task is to control the size of the values computed by the
threads.  To this end, we propose a suitable notion of
quasi-interpretation (cf.~\cite{BMM01,Amadio02,Amadio04}).

\begin{definition}[assignment]\label{assignment}
  Given a program, an {\em assignment} $q$ associates with
  constructors and function symbols, functions over the non-negative
  reals $\Real$ such that:
  \begin{description}
  \item[{\rm (1)}] If $\s{c}$ is a constant then $q_{\s{c}}$ is the
    constant $0$.
  
  \item[{\rm (2)}] If $\s{c}$ is a constructor with arity $n\geq 1$
    then $q_{\s{c}}$ is a function in $(\Real )^{n} \to \Real$ such
    that $q_{\s{c}}(x_1, \ldots, x_n) = d + \Sigma_{i\in 1..n} x_i$,
    for some $d\geq 1$.
  
  \item[{\rm (3)}] If $f$ is a function (name) with arity $n$ then
    $q_f : (\Real )^{n} \to \Real$ is monotonic and for all $i \in
    1..n$ we have $q_f(x_1, \ldots, x_n) \geq x_i$.
  \end{description}
\end{definition}

An assignment $q$ is extended to all expressions $e$ as follows,
giving a function expression $q_e$ with variables in $\w{Var}(e)$:
\[
\begin{array}{c@{\quad}c@{\quad}c}
  q_x = x~,

  &q_{\s{c}(e_1,\ldots,e_n)} =  q_{\s{c}}(q_{e_{1}},\ldots,q_{e_{n}})~,

  &q_{f(e_1,\ldots,e_n)} =  q_{f}(q_{e_{1}},\ldots,q_{e_{n}})~.
\end{array}
\]

Here $q_x$ is the identity function and, e.g.,
$q_{\s{c}}(q_{e_{1}},\ldots,q_{e_{n}})$ is the functional composition
of the function $q_{\s{c}}$ with the functions $q_{e_{1}},\ldots,
q_{e_{n}}$. It is easy to check that there exists a constant
$\delta_q$ depending on the assignment $q$ such that for all values
$v$ we have $|v|\leq q_v \leq \delta_q \cdot |v|$.  Thus the
quasi-interpretation of a value is always proportional to its size.

\begin{definition}[quasi-interpretation]
  An assignment is a {\em quasi-interpreta\-tion}, if for all
  constraints associated with the system of the shape
  $f(\vc{p})\succ_i e$, with $i\in \set{0,1}$, the inequality
  $q_{f(\vc{p})} \ \geq\ q_e$ holds over the non-negative reals.
\end{definition}

Quasi-interpretations are designed so as to provide a bound on the
size of the computed values as a function of the size of the input
data. In the following, we assume given a suitable
quasi-interpretation, $q$, for the system under investigation.

\begin{example}\label{max-value-ex-qint}
  With reference to Examples~\ref{example:exponent}
  and~\ref{max-value-ex}, the following assignment is a
  quasi-interpretation (the parameter $i$ corresponds to the label of
  the \w{read} instruction in the body of $f$). We give no
  quasi-interpretations for the function $\w{exp}$ because it fails
  the read once condition:
  \[{\small
  \begin{array}{l}
    q_{\s{nil}} = q_{\s{z}} = 0\ ,\quad
    q_{\s{s}}(x) = x+1\ ,\quad q_{\s{cons}}(x,l) = x+l+1\ ,
    \quad q_{\w{dble}}(x) = 2\cdot x\ ,\\
    q_{\hatt{f}}(x, i)  =   x+i\ ,\qquad q_{\hatt{f}_{1}}(x) = x\ ,\qquad
    q_{\w{maxl}}(x,y) = q_{max}(x,y) = \w{max}(x,y)~.
  \end{array}}
  \]
\end{example}

One can show~\cite{Amadio02,Amadio04} that in the purely functional
fragment of our language every value $v$ computed during the
evaluation of an expression $f(v_1,\ldots,v_n)$ satisfies the
following condition:
\begin{equation}\label{qint-prop}
|v| \ \leq\ q_v \ \leq\ q_{f(v_1,\ldots,v_n)} \ =\ 
q_{f}(q_{v_1},\ldots,q_{v_n}) \ \leq\ q_f(\delta_q \cdot |v_1|,
\ldots,\delta_q \cdot |v_n|)~.
\end{equation}
We generalise this result to threads as follows.

\begin{theorem}[bound on the size of the values]\label{thread-bound}
  Given a system of synchronous threads $B$, suppose that at the
  beginning of the instant $B_{1}(i)=f(\vc{v})$ for some thread index
  $i$. Then the size of the values computed by the thread $i$ during
  an instant is bounded by $q_{\hatt{f}(\vc{v},\vc{u})}$ where
  $\vc{u}$ are the values contained in the registers at the time they
  are read by the thread (or some constant value, if they are not read
  at all).
\end{theorem}

Theorem \ref{thread-bound} is proven by showing that
quasi-interpretations satisfy a suitable invariant. In the following
corollary, we note that it is possible to express a bound on the size
of the computed values which depends only on the size of the
parameters at the beginning of the instant. This is possible because
the number of reads a system may perform in an instant is bounded by a
constant.

\begin{corollary}\label{sys-bound}
  Let $B$ be a system with $m$ distinct \w{read} instructions and $n$
  threads.  Suppose $B_1(i)=f_i(\vc{v}_i)$ for $i\in \Z_n$. Let $c$ be
  a bound of the size of the largest parameter of the functions
  $f_{i}$ and the largest default value of the registers. Suppose $h$
  is a function bounding all the quasi-interpretations, that is, for
  all the functions $\hatt{f}_{i}$ we have $h(x)\geq
  \smash[b]{q_{\hatt{f}_{i}}(x,\ldots,x)}$ over the non-negative
  reals. Then the size of the values computed by the system $B$ during
  an instant is bounded by $h^{n\cdot m + 1}(c)$.
\end{corollary}

\begin{example}
  The $n\cdot m$ iterations of the function $h$ predicted by
  Corollary~\ref{sys-bound} correspond to a tight bound, as shown by
  the following example. We assume $n$ threads and one register, $r$,
  of type \w{nat} with default value \s{z}. The control of each thread
  is described as follows:
  \[{\small
  \begin{array}{ll}
    f(x_0)\ = &\w{read}\, \s{r} \,\w{with}\, x_1 \Arrow 
    \s{r} := \w{dble}(\w{max}(x_1,x_0)).\\
    &\quad \w{read}\, \s{r} \,\w{with}\, x_2  \Arrow 
    \s{r} := \w{dble}(x_{2}).\\
    & \qquad \quad \ldots\ldots \\
    & \qquad 
    \w{read}\, \s{r} \,\w{with}\, x_m \Arrow 
    \s{r} := \w{dble}(x_m).\w{next}.{f(\w{dble}(x_m))} ~.
  \end{array}}
  \]
  
  For this system we have $c \geq |x_0|$ and $h(x) = q_\w{dble}(x) =
  2\cdot x$.  It is easy to show that, at the end of an instant, there
  have been $n\cdot m$ assignments to the register \s{r} ($m$ for
  every thread in the system) and that the value stored in \s{r} is
  $\w{dble}^{n\cdot m}(x_0)$ of size $2^{n\cdot m}\cdot |x_0|$.
\end{example}

\subsection{Combining Termination and 
  Quasi-interpretations}\label{size-termination}
To bound the space needed for the execution of a system
during an instant we also need to bound the number of nested recursive
calls, i.e. the number of frames that can be found on the stack (a
precise definition of frame is given in the following
Section~\ref{vm}). Unfortunately, quasi-interpretations provide a
bound on the size of the frames but not on their number (at least not
in a direct implementation that does not rely on memoization).  One
way to cope with this problem is to combine quasi-interpretations with
various families of reduction orders~\cite{Marion00,BMM01}. In the
following, we provide an example of this approach based on {\em
  recursive path orders} which is a widely used and fully mechanizable
technique to prove termination~\cite{BaaderNipkow98}.

\begin{definition}\label{def-LPO-termination}
  We say that a system terminates by LPO, if the reduction order
  associated with the system is a recursive path order where: {(1)} 
  symbols are ordered so that function symbols are always bigger than
  constructor symbols and two  distinct constructor symbols are
  incomparable; {(2)} the arguments of function symbols are compared 
  with respect to the lexicographic order and those of 
  constructor symbols with respect to the product order.
\end{definition}

Note that because of the hypotheses on constructors, this is actually
a special case of the lexicographic path order. For the sake of brevity,
we still refer to it as LPO.
\begin{definition}
  We say that a system admits a polynomial quasi-interpretation if it
  has a quasi-interpretation where all functions are bounded by a
  polynomial.
\end{definition}

The following property is a central result of this paper.
\begin{theorem}\label{pspace-bound}
  If a system $B$ terminates by LPO and admits a polynomial
  quasi-interpretation then the computation of the system in an
  instant runs in space polynomial in the size of the parameters of
  the threads at the beginning of the instant.
\end{theorem}

The proof of Theorem \ref{pspace-bound} is based on
Corollary~\ref{sys-bound} that provides a polynomial bound on the size
of the computed values and on an analysis of nested calls in the LPO
order that can be found in~\cite{BMM01}. The point is that the depth
of such nested calls is polynomial in the size of the values and that
this allows to effectively compute a polynomial bounding the space
necessary for the execution of the system. 

\begin{example}\label{max-value-ex-pspace}
  We can check that the order used in Example~\ref{max-value-ex} for
  the functions $\hatt{f}, \hatt{f_1}, \w{max}$ and $\w{maxl}$ is
  indeed a LPO. Moreover, from the quasi-inter\-pretation given in
  Example~\ref{max-value-ex-qint}, we can deduce that the function
  $h(x)$ has the shape $a\!\cdot\!x+b$ (it is affine). In practice,
  many useful functions admit quasi-interpretations bound by an affine
  function such as the max-plus polynomials considered
  in~\cite{Amadio02,Amadio04}.
\end{example}

The combination of LPO and polynomial quasi-interpretation actually
provides a characterisation of PSPACE. In order to get to PTIME a
further restriction has to be imposed. Among several possibilities, we
select one proposed in \cite{BMM04}. We say that the system terminates
by {\em linear} LPO if it terminates by LPO as in definition
\ref{def-LPO-termination} and moreover if in all the constraints
$f(\vc{p})\succ_0 e$ or $f^+(\vc{p})\succ_0 g^+(\vc{e})$ of index 0
there is at most one function symbol on the right hand side which has
the same priority as the (unique) function symbol on the left-hand
side. For instance, the Example~\ref{max-value-ex} falls in this
case. In op. cit., it is shown by a simple counting argument that
the number of calls a function may generate is polynomial
in the size of its arguments. One can then restate 
theorem \ref{pspace-bound} by replacing LPO with linear LPO and 
PSPACE with PTIME.\\
We stress that these results are of a {\em constructive} nature,
thus beyond proving that a system `runs in PSPACE (or PTIME)', 
we can extract a definite polynomial that bounds the size needed 
to run a system during an instant. In general, the bounds are
rather rough and should be regarded as providing
a {\em qualitative} rather than {\em quantitative} information. \\
In the purely functional framework, M.~Hofmann \cite{Hofmann02} has
explored the situation where a program is {\em non-size increasing}
which means that the size of all intermediate results is bounded
by the size of the input. Transferring this concept to a system of
threads is attractive because it would allow to predict the behaviour
of the system for arbitrarily many instants.  However, this is 
problematic. For instance, consider again example
\ref{max-value-ex-pspace}.  By Theorem~\ref{pspace-bound}, we can
prove that the computation of a system running the behaviour $f(x_0)$
in an instant requires a space polynomial in the size of $x_0$. Note
that the parameter of $f$ is the largest value received so far in the
register \s{i}.  Clearly, bounding the value of this parameter for
arbitrarily many instants requires a {\em global} analysis of the
system which goes against our wish to produce a {\em compositional}
analysis in the sense explained in the Introduction.  An alternative
approach which remains to be explored could be to develop linguistic
tools and a programming discipline that allow each thread to control
{\em locally} the size of its parameters.

\section{A Virtual Machine}\label{vm}
We describe a simple virtual machine for our language thus providing a
concrete intuition for the data structures required for the execution
of the programs and the scheduler.\\
Our motivations for introducing a low-level model of execution for
synchronous threads are twofold: (i) it offers a simple formal
definition for the space needed for the execution of an instant (just
take the maximal size of a machine configuration), and (ii) it
explains some of the elaborate mechanisms occurring during the
execution, like the synchronisation with the \w{read} instruction and
the detection of the end of an instant.  A further motivation which is
elaborated in Section~\ref{cfa-rev} is the possibility to carry on the
static analyses for resource control at bytecode level.  The interest
of bytecode verification is now well understood, and we refer the
reader to~\cite{MWCG99,Necula97}.

\subsection{Data Structures}
We suppose given the code for all the threads running in a system
together with a set of types and {\em constructor names} and a
disjoint set of {\em function names}. A function name $f$ will
also denote the sequence of instructions of the associated code:
$f[i]$ stands for the $i^\w{th}$ instruction in the (compiled) code of
$f$ and $|f|$ stands for the number of instructions.\\
The configuration of the machine is composed of a \emph{store} $s$,
that maps registers to their current values, a sequence of records
describing the state of each thread in the system, and three local
registers owned by the scheduler whose role will become clear in
Section~\ref{scheduler}.\\
A thread identifier, $t$, is simply an index in $\Z_n$. The state of a
thread $t$ is a pair $(\w{st}_t, M_t)$ where $\w{st}_t$ is a
\emph{status} and $M_t$ is the \emph{memory} of the thread. A
\emph{memory} $M$ is a sequence of frames, and a \emph{frame} is a
triple $(f, \w{pc}, \ell)$ composed of a function name, the value of
the program counter (a natural number in $1..|f|$), and a \emph{stack}
of values $\ell = v_1 \cdots v_k$. We denote with $|\ell|$ the number
of values in the stack.  The status of a thread is defined as in the
source language, except for the status $W$ which is refined into
$W(j,n)$ where: $j$ is the index where to jump at the next instant if
the thread does not resume in the current instant, and $n$ is the
(logical) time at which the thread is suspended (cf.
Section~\ref{scheduler}).

\subsection{Instructions}
The set of \emph{instructions} of the virtual machine together with
their operational meaning is described in
Table~\ref{bytecode-instructions}. All instructions operate on the
frame of the current thread $t$ and the memory $M_t$ --- the only
instructions that depend on or affect the store are \texttt{read} and
\texttt{write}.  For every segment of bytecode, we require that the
last instruction is either $\Return$, $\Stop$ or $\Tcall$ and that the
jump index $j$ in the instructions $\Branch \ \s{c} \ j$ and $\Wait \ 
j$ is within the segment.

\begin{table}
  \caption{Bytecode instructions}\label{bytecode-instructions}
  {\small
    \[
    \begin{array}{l|@{\ }l@{}}
      
      f[\pc] 
      &\mbox{\ Current memory}  
      \qquad\qquad\qquad \mbox{\ Following memory}\\ \hline
      
      \Load\ k
      &M \concat ( f , \pc , \ell\cdot v \cdot \ell') 
      ~\act{}~ M \concat ( f , \pc + 1 ,  \ell \cdot v \cdot \ell'\cdot v),
      ~|\ell|=k-1
      \\

      \Branch\ \s{c}\ j 
      &M \concat (f, \pc, \ell \concat \s{c}(v_1, \ldots, v_n) ) 
      ~\act{}~   M \concat (f ,\pc+1, \ell \concat v_1\cdots v_n) \\ 
      
      \Branch\ \s{c}\ j 
      &M \concat (f, \pc, \ell\concat \s{d}(\ldots))
      ~\act{}~ M \concat (f, j, \ell\concat \s{d}(\ldots))
      ~~\s{c} \neq \s{d} \\
  
      \Build\ \s{c}\ n 
      &M \concat (f, \pc, \ell\concat v_1 \cdots v_n)
      ~\act{}~ M\concat (f, \pc + 1, \ell \concat \s{c}(v_1, \ldots,
      v_n) ) \\
  
      \Call\ g\ n 
      &M \concat (f, \pc, \ell \concat v_1 \cdots v_n )
      ~\act{}~ M\concat (f, \pc, \ell \concat v_1 \cdots v_n) \concat (g, 1, v_1\cdots v_n) \\
      
      \Tcall\ g\ n 
      &M \concat (f, \pc, \ell \concat v_1 \cdots v_n)
      ~\act{}~ M\concat (g, 1, v_1\cdots v_n) \\

      \Return 
      &M \concat (g, \pc', \ell' \cdot \vc{v'}) \concat
      (f, \pc, \ell \concat v) 
      ~\act{}~ M \concat (g, \pc' + 1,  \ell' \concat v),~\w{ar}(f)=|\vc{v'}| \\

      \Read \ \s{r}
      &(M\concat (f, \pc, \ell),s) 
      ~\act{}~ (M\concat (f, \pc+1, \ell \concat s(\s{r})),s) \\

      \Read \ k
      &(M\concat (f, \pc, \ell \cdot \s{r} \cdot \ell'),s)
      ~\act{}~ (M\concat (f, \pc+1, \ell \cdot \s{r} \cdot
      \ell' \concat s(\s{r})),s),~ |\ell|=k-1  \\
      
      \Write  \ \s{r}
      &(M\concat (f, \pc, \ell \concat v),s)
      ~\act{}~ (M\concat (f, \pc+1, \ell),s[v/\s{r}]) \\

      \Write  \ k
      &(M\concat (f, \pc, \ell \cdot \s{r} \cdot \ell' \concat v),s)
      \act{} (M\concat (f, \pc+1, \ell \cdot \s{r} \cdot \ell'),s[v/\s{r}]),
      |\ell|=k-1\\
      
      \Stop
      &M \concat (f, \pc, \ell) 
      ~\act{S}~ \Error \\
      
      \Yield
      &M\concat (f, \pc, \ell) 
      ~\act{R}~ M\concat (f, \pc+1, \ell) \\
      
      \Next
      &M\concat (f, \pc, \ell) 
      ~\act{N}~ M\concat (f, \pc+1, \ell) \\
      
      \Wait \ j
      &M\concat (f, \pc, \ell\cdot v) 
      ~\act{W}~ M\concat (f, j, \ell)
    \end{array} \]}
\end{table}

\begin{table}
\caption{An implementation of the
  scheduler}\label{scheduler-implementation}
{\small
  \[
  \begin{array}{@{\quad}c@{\quad}}
\\
    \begin{array}{l@{\ }l}
      \s{for}\,  t \,\s{in}\, \Z_n \,\s{do}\, \{\ \w{st}_t :=R; \ \}
      &\text{(initialisation)}\\
      s := s_o; \s{tid} := \s{time} := \s{wtime} := 0;
      &\text{(the initial thread is of index $0$)} \\
      \s{while}\ (\s{tid} \in \Z_n)
      \ \{      &\text{(loop until all threads are blocked)}\\
      \ \s{if}\, I_{\s{tid}} = (\s{write}\ \_\,) \,\s{then}\, \s{wtime}
      := \s{time}; &\text{(record store modified)} \\
      \ \s{if}\, I_{\s{tid}} = (\s{wait}\ j\,)\\
      \ \s{then}\, \w{st}_{\s{tid}} :=
      W(\pc_{\s{tid}}+1,\s{time}); &\text{(save continuation for next
        instant)}\\
      \ X:=\w{run}(\s{tid}); &\text{(run current
        thread)}\\
      \ \s{if}\ X\neq \epsilon \ \s{then} \ \{\\
      \ \quad \s{if}\, X \neq W \, \s{then}\, 
      \w{st}_{\s{tid}}:= X; &\text{(update thread status)} \\
      \ \quad  \s{tid} := \cl{N}(\s{tid}, \w{st});
      &\text{(compute index of next active thread)} \\
      \ \quad \s{if}\, \s{tid} \in \Z_ n 
      &\text{(test whether all threads are blocked)} \\
      \ \quad\s{then}\, \{\
      \w{st}_{\s{tid}}:=R;\, \s{time}:=\s{time}+1; \}
      &\text{(if not, prepare next thread to run)}\\
      \ \quad \s{else}\, \{\ 
      s := s_o;\, \s{wtime} := \s{time}; &\text{(else, initialisation of the
        new instant)}\\
      \ \qquad \quad \s{tid} := \cl{N}(0, \w{st}); &\text{(select thread
        to run, starting from $0$)}\\
      \ \qquad \quad \s{forall}\, i \,\s{in}\, \Z_n \,\s{do}\ \{\\
      \ \qquad\qquad \s{if}\, \w{st}_i = W(j,\_) \,\s{then}\, \w{pc}_i := j ;\\
      \ \qquad\qquad \s{if}\, 
      \w{st}_i \neq S \,\s{then}\, \w{st}_i := R; \ \}\, \}\, \}\\[1em]
    \end{array}~\\
    
    \textsc{Conditions on}~\cl{N}:\\
    
    \begin{array}{ll}
      \mbox{If }\cl{N}(\s{tid},\w{st})=k\in \Z_n
      &\mbox{then }\w{st}_k=R \mbox{ or }(\w{st}_k=W(j,n) \mbox{ and }
      n<\s{wtime}) \\
    
      \mbox{If }\cl{N}(\s{tid},\w{st})\notin \Z_n
      &\mbox{then }\forall k \in \Z_n \, 
      \begin{array}[t]{l}
        (\w{st}_k\neq R \mbox{ and }\\
        \qquad (\w{st}_k=W(j,n) \mbox{ implies }n\geq \s{wtime}) )\\[1em]
      \end{array}
    \end{array}
  \end{array}
  \]}
\end{table}

\subsection{Scheduler}\label{scheduler}
In Table~\ref{scheduler-implementation} we describe a simple
implementation of the scheduler. The scheduler owns three registers:
(1) $\s{tid}$ that stores the identity of the current thread, (2)
\s{time} for the current time, and (3) \s{wtime} for the last time the
store was modified. The notion of time here is of a logical nature:
time passes whenever the scheduler transfers control to a new thread.
Like in the source language, $s_o$ denotes the store at the beginning
of each instant.\\
The \emph{scheduler} triggers the execution of the current instruction
of the current thread, whose index is stored in $\s{tid}$, with a call
to $\w{run}(\s{tid})$. The call returns the label $X$ associated with
the instruction in Table~\ref{bytecode-instructions}.  By convention,
take $X=\epsilon$ when no label is displayed. If $X\neq\epsilon$ then
the scheduler must take some action. Assume $\s{tid}$ stores the
thread index $t$. We denote $\pc_{\s{tid}}$ the program counter of the
top frame $(f, \w{pc}_{t}, \ell)$ in $M_t$, if any, $I_{\s{tid}}$ the
instruction $f[\w{pc}_t]$ (the current instruction in the thread) and
$\w{st}_{\s{tid}}$ the state $\w{st}_t$ of the thread.  Let us explain
the role of the status $W(j,n)$ and of the registers \s{time} and
\s{wtime}. We assume that a thread waiting for a condition to hold can
check the condition without modifying the store.  Then a thread
waiting since time $m$ may pass the condition only if the store has
been modified at a time $n$ with $m<n$.  Otherwise, there is no point
in passing the control to it\footnote{Of course, this condition can be
  refined by recording the register on which the thread is waiting,
  the shape of the expected value,$\ldots$}. With this data structure
we also have a simple method to detect the end of an instant, it
arises when no thread is in the running status and all waiting threads
were interrupted after the last store modification occurred.\\
In models based on preemptive threads, it is difficult to foresee the
behaviour of the scheduler which might depend on timing information
not available in the model. For this reason and in spite of the fact
that most schedulers are deterministic, the scheduler is often
modelled as a non-deterministic process. In cooperative threads, as
illustrated here, the interrupt points are explicit in the program and
it is possible to think of the scheduler as a deterministic process.
Then the resulting model is deterministic and this fact considerably
simplifies its programming, debugging, and analysis.

\begin{table}
\caption{Compilation of source code to bytecode}\label{compilation}
{\small
  \[\begin{array}[c]{@{\quad}c@{\quad}}
  \mbox{\normalsize\textsc{Compilation of expression bodies:}}\\[1em]
  \begin{array}{ll}
    C(e,\eta)       
    &=\ C'(e,\eta)\cdot \Return\\
      
    C\left(
      \begin{array}{@{}c@{}}
        \w{match}~x~\w{with}~\s{c}(\vc{y})\\
        \w{then}~{\w{eb}_{1}}~\w{else}~{\w{eb}_{2}}     
      \end{array}, \eta \right)
    &=\ \left\{ \begin{array}{@{}l}
        (\Branch \ \s{c} \ j) \cdot C(\w{eb}_{1}, \eta'\cdot \vc{y})
        \cdot \mbox{ if }\eta=\eta' \cdot x\\
        \quad (j: C(\w{eb}_{2}, \eta))
         \\
        (\Load \ i(x,\eta)) \cdot (\Branch \ \s{c} \ j) \cdot \mbox{ o.w.}\\
        \quad C(\w{eb}_{1}, \eta \cdot \vc{y}) \cdot (j: C(\w{eb}_{2}, \eta\cdot x))
      \end{array}\right. \\[1em]
  \end{array}\\

  \mbox{\normalsize\textsc{Auxiliary compilation of expressions:}}\\[1em]
  \begin{array}{ll}
    C'(x,\eta)      
    &=\ (\Load \ i(x,\eta)) \\
    
    
    C'(\s{c}(e_{1},\ldots,e_{n}),\eta) 
    &=\ C'(e_{1},\eta) \cdot \ldots \cdot C'(e_{n},\eta) \cdot (\Build \ \s{c} \ n) \\
    
    C'(f(e_{1},\ldots,e_{n}),\eta) 
    &=\ C'(e_{1},\eta) \cdot \ldots \cdot C'(e_{n},\eta)\cdot (\Call \ f \ n) \\[1em]
  \end{array}\\

  \mbox{\normalsize\textsc{Compilation of behaviours:}}\\[1em]
  \begin{array}{ll}
    C(\w{stop},\eta)  
    &=\  \Stop \\
    
    C(f(e_{1},\ldots,e_{n}),\eta)
    &=\  C'(e_{1},\eta) \cdots C'(e_{n},\eta) \cdot (\Tcall \ f \ n)\\
    
    C(\w{yield}.b,\eta)
    &=\  \Yield \cdot C(b,\eta) \\
    
    C(\w{next}.f(\vc{e}),\eta)
    &=\  \Next \cdot C(f(\vc{e}),\eta) \\
    
    C(\varrho := e.b,\eta)
    &=\  C'(e,\eta) \cdot  (\Write \ i(\varrho,\eta)) \cdot C(b,\eta) \\
    
    C\left(
      \begin{array}{@{}c@{}}
        \w{match}~x~\w{with}~\s{c}(\vc{y})\\
        \w{then}~{b_{1}}~\w{else}~{b_{2}}     
      \end{array}, \eta \right)
    &=\  
    \left\{ \begin{array}{@{}l}
        (\Branch \ \s{c} \ j) \cdot C(b_{1}, \eta'\cdot \vc{y})
        \cdot \mbox{ if }\eta=\eta'\cdot x\\
        \quad (j: C(b_{2}, \eta))
        \\
        (\Load \ i(x,\eta)) \cdot (\Branch \ \s{c} \ j)\cdot \mbox{ o.w.}\\
        \quad C(b_{1}, \eta \cdot \vc{y}) \cdot (j: C(b_{2}, \eta\cdot x))
      \end{array}\right. \\

C\left(
      \begin{array}{@{}c@{}}
        \rmatch{\varrho} \cdots \mid \s{c}_{\ell}(\vc{y}_{\ell}) \Arrow b_{\ell} \mid  \\
        ~~~~~~~~~~~~~\cdots y_k \Arrow b_k \cdots
      \end{array} , \eta \right)
    &=\  
    \left(\begin{array}{@{}l}
        j_{0} : (\Read \ i(\varrho,\eta)) \cdot \ldots \cdot\\
        j_{\ell} : (\Branch \ \s{c}_{\ell} \ j_{\ell+1}) \cdot
        C(b_{\ell},\eta \cdot \vc{y}_{\ell})\cdot \\
        j_{\ell+1} : \cdots j_{k} : C(b_{k},\eta\cdot y_k)
      \end{array}\right)\\
    
    C\left(
      \begin{array}{@{}c@{}}
        \rmatch{\varrho} \cdots \mid \s{c}_{\ell}(\vc{y}_{\ell}) \Arrow b_{\ell} \mid  \\
        ~~~~~~~~~~~~~\cdots \mid [\_] \Arrow g(\vc{e})
      \end{array} , \eta \right)
    &=\  
    \left(\begin{array}{@{}l}
        j_{0} : (\Read \ i(\varrho,\eta)) \cdot \ldots \cdot\\
        j_{\ell} : (\Branch \ \s{c}_{\ell} \ j_{\ell+1}) \cdot
        C(b_{\ell},\eta \cdot \vc{y}_{\ell}) \cdot\\
        j_{\ell+1} : \cdots j_{n}:(\Wait\ j_{0}) \cdot C(g(\vc{e}),\eta)\\[.2em]
      \end{array}\right)\\[1em]
  \end{array}\\[1em]
\end{array}\]}
\end{table}

\subsection{Compilation}
In Table~\ref{compilation}, we describe a possible compilation of the
intermediate language into bytecode.  We denote with $\eta$ a sequence
of variables.  If $x$ is a variable and $\eta$ a sequence then
$i(x,\eta)$ is the index of the rightmost occurrence of $x$ in $\eta$.
For instance, $i(x,x\cdot y \cdot x) =3$. By convention,
$i(\s{r},\eta)=\s{r}$ if $\s{r}$ is a register constant.  We also use
the notation $j:C(\w{be},\eta)$ to indicate that $j$ is the position
of the first instruction of $C(\w{be},\eta)$. This is just a
convenient notation since, in practice, the position can be computed
explicitly.  With every function definition
$f(x_{1},\ldots,x_{n})=\w{be}$ we associate the bytecode
$C(\w{be},x_{1}\cdots x_{n})$.
\begin{example}[compiled code]
  We show below the result of the compilation of the function
  \w{alarm} in Example~\ref{alarm-function}:
  \[{\small
    \begin{array}{l@{\ :\ }l@{\qquad}l@{\ :\ }l@{\qquad}l@{\ :\ }l}
      1 &\Branch \ \s{s}\ 12           &6 &\Load \ 1
      &11 &\Tcall \ \w{alarm} \ 2 \\
      2 &\Read \ \s{sig}           &7 &\Tcall \ \w{alarm} \ 2 
      &12 &\Build \ \s{prst} \ 0   \\
      3 &\Branch \ \s{prst}\ 8
      &8 &\Wait\ 2
      &13 &\Write \ \s{ring}\\
      4 &\Next                     &9 &\Load \ 1   
      &14 &\Stop \\
      5 &\Load\ 1                   &10 &\Load \ 2
    \end{array}}
    \]
\end{example}

\subsection{Control Flow Analysis Revisited}\label{cfa-rev}
As a first step towards control flow analysis, we analyse the {\em
  flow graph} of the bytecode generated.

\begin{definition}[flow graph]
  The flow graph of a system is a directed graph whose nodes are pairs
  $(f,i)$ where $f$ is a function name in the program and $i$ is an
  instruction index, $1\leq i \leq |f|$, and whose edges are
  classified as follows:
  \begin{description}
    
  \item[Successor:] An edge $((f,i),(f,i+1))$ if $f[i]$ is a $\Load$,
    $\Branch$, $\Build$, $\Call$, $\Read$, $\Write$, or $\Yield$
    instruction.
    
  \item[Branch:] An edge $((f,i), (f,j))$ if $f[i]=\Branch \ \s{c} \ 
    j$.
    
  \item[Wait:] An edge $((f,i), (f,j))$ if $f[i]=\Wait \ j$.
    
  \item[Next:] An edge $((f,i), (f,i+1))$ if $f[i]$ is a $\Wait$ or
    $\Next$ instruction.
    
  \item[Call:] An edge $((f,i), (g,1))$ if $f[i]=\Call \ g \ n$ or
    $f[i]=\Tcall \ g \ n$.

  \end{description}
\end{definition}

The following is easily checked by inspecting the compilation
function. Properties \textbf{Tree} and \textbf{Read-Wait} entail that
the only cycles in the flow graph of a function correspond to the
compilation of a \w{read} instruction. Property \textbf{Next} follows
from the fact that, in a behaviour, an instruction \w{next} is always
followed by a function call $f(\vc{e})$. Property \textbf{Read-Once}
is a transposition of the read once condition (Section~\ref{cfa}) at
the level of the bytecode.
\begin{proposition}\label{flow-graph-prop}
  The flow graph associated with the compilation of a well-formed
  system satisfies the following properties:
  \begin{description}
  
  \item[Tree:] Let $G'$ be the flow graph without wait and call edges.
    Let $G'_f$ be the full subgraph of $G'$ whose nodes have the shape
    $(f,i)$.  Then $G'_f$ is a {\em tree} with root $(f,1)$.
  
  \item[Read-Wait:] If $f[i]=\Wait \ j$ then $f[j] = \Read \ r$ and
    there is a unique path from $(f,j)$ to $(f,i)$ and in this path,
    every node corresponds to a $\Branch$ instruction.
  
  \item[Next:] Let $G'$ be the flow graph without call edges.  If
    $((f,i), (f,i+1))$ is a next edge then for all nodes $(f,j)$
    accessible from $(f,i+1)$, $f[j]$ is not a $\Read$ instruction.
  
  \item[Read-Once:] Let $G'$ be the flow graph without wait edges and
    next edges. If the source code satisfies the read once condition
    then there is no loop in $G'$ that goes through a node $(f,i)$
    such that $f[i]$ is a $\Read$ instruction.
  \end{description}
\end{proposition}

In~\cite{ACDJ04}, we have presented a method to perform resource
control verifications at bytecode level. This work is just concerned
with the functional fragment of our model. Here, we outline its
generalisation to the full model. The main problem is to reconstruct a
symbolic representation of the values allocated on the stack. Once
this is done, it is rather straightforward to formulate the
constraints for the resource control. We give first an informal
description of the method.
\begin{enumerate}
  
\item For every segment $f$ of bytecode instructions with, say, formal
  parameters $x_1,\ldots, x_n$ and for every instruction $i$ in the
  segment, we compute a sequence of expressions $e_1 \cdots e_m$ and a
  substitution $\sigma$.
  
\item The expressions $(e_i)_{i \in 1..m}$ are related to the formal
  parameters via the substitution $\sigma$. More precisely, the
  variables in the expressions are contained in $\sigma
  x_1,\ldots,\sigma x_n$ and the latter forms a linear pattern.
  
\item Next, let us look at the intended usage of the formal
  expressions.  Suppose at run time the function $f$ is called with
  actual parameters $u_1,\ldots,u_n$ and suppose that following this
  call, the control reaches instruction $i$ with a stack $\ell$. Then
  we would like that:
  \begin{itemize}
  
  \item The values $u_1,\ldots,u_n$ match the pattern $\sigma
    x_{1},\ldots,\sigma x_n$ via some substitution $\rho$.
  
  \item The stack $\ell$ contains exactly $m$ values $v_1,\ldots,v_m$
    whose types are the ones of $e_1,\ldots,e_m$, respectively.
  
  \item Moreover $\rho (e_i)$ is an {\em over-approximation} (w.r.t.
    size and/or termination) of the value $v_i$, for $i=1,\ldots, m$.
    In particular, if $e_i$ is a pattern, we want that
    $\rho(e_i)=v_i$.
  \end{itemize} 
\end{enumerate}

We now describe precisely the generation of the expressions and the
substitutions. This computation is called \emph{shape analysis}
in~\cite{ACDJ04}. For every function $f$ and index $i$ such that
$f[i]$ is a $\Read$ instruction we assume a fresh variable $x_{f,i}$.
Given a total order on the function symbols, such variables can be
totally ordered with respect to the index $(f,i)$.  Moreover, for
every index $i$ in the code of $f$, we assume a countable set
$x_{i,j}$ of distinct variables.\\
We assume that the bytecode comes with annotations assigning a
suitable type to every constructor, register, and function symbol.
With every function symbol $f$ of type $\vc{t} \arrow \w{beh}$, comes
a fresh function symbol $\hatt{f}$ of type $\vc{t}, \vc{t'}\arrow
\w{beh}$ so that $|\vc{t'}|$ is the number of read instructions
accessible from $f$ within an instant.  Then, as in the definition of
control points (Section~\ref{sec:control-points}), the extra arguments
in $\hatt{f}$ corresponds to the values read in the registers within
an instant. The order is chosen according to the order of the
variables associated with the $\Read$ instructions.\\
In the shape analysis, we will consider well-typed expressions
obtained by composition of such fresh variables with function symbols,
constructors, and registers. In order to make explicit the type of a
variable $x$ we will write $x^t$.\\
For every function $f$, the shape analysis computes a vector
$\vc{\sigma}= {\sigma}_{1}, \ldots, {\sigma}_{|f|}$ of
substitutions and a vector $\vc{E}= E_1,\ldots,E_{|f|}$ of sequences
of well-typed expressions. We let $\vc{E}_i$ and $\vc{\sigma}_i$
denote the sequence $E_i$ and the substitution $\sigma_i$ respectively
(the $i^{\w{th}}$ element in the vector), and $\vc{E}_i[k]$ the
$\smash[t]{k^{\w{th}}}$ element in $\vc{E}_{i}$.  We also let
$h_{i}=|\vc{E}_{i}|$ be the length of the $i^{\w{th}}$ sequence. We
assume $\vc{\sigma}_{1}=\w{id}$ and $\vc{E}_1 =
x_{1,1}^{t_{1}}\cdots x_{1,n}^{t_{n}}$, if
$f:t_1,\ldots,t_n\arrow \beta$ is a function of arity $n$.\\
The main case is the $\Branch$ instruction:
{\small
  \[
  \begin{array}{l|@{\quad}l}

    \ f[i]=           &
    \ \mbox{Conditions} \\\hline
    
    \Branch \ \s{c} \ j
    \ &\s{c}:\vc{t} \arrow t,\ 
    \vc{E}_{i}=E\cdot e,\  
    e:t,\\

    &\mbox{ and either }  
    e=\s{c}(\vc{e}),\ 
    \vc{\sigma}_{i+1} = \vc{\sigma}_{i},\
    \vc{E}_{i+1}=E \cdot \vc{e}\\

    &\mbox{ or } e=\s{d}(\vc{e}),\ 
    \s{c}\neq \s{d},\  
    \vc{\sigma}_{j} = \vc{\sigma}_{i},\ 
    \vc{E}_{j}=\vc{E}_{i}\\

    &\mbox{ or } e=x^{t},\ 
    \vc{\sigma}_j = \vc{\sigma}_i,\
    \vc{E}_{j} = \vc{E}_i,\
    \sigma' = 
    [\s{c}(x_{i+1,{h}_{i}}^{t_{1}},\ldots,x_{i+1,{h}_{i+1}}^{t_{n}})/x], \\

    &\quad \vc{\sigma}_{i+1} = \sigma' \comp \vc{\sigma}_i,\ 
    \vc{E}_{i+1} = \sigma'(E) 
    \concat x_{i+1, {h}_i} \cdots x_{i+1, {h}_{i+1}}~. 
  \end{array}
  \]}

The constraints for the remaining instructions are given in
Table~\ref{shape-an}, where it is assumed that $\vc{\sigma}_{i+1}
= \vc{\sigma}_i$ except for the instructions $\Tcall$ and
$\Return$ (that have no direct successors in the code of the
function).

\begin{table}
\caption{Shape analysis at bytecode level}\label{shape-an}
{\small
  \[
  \begin{array}{l|@{\quad}l}
    
    \ f[i]=           &\ \mbox{Conditions} \\\hline

    \Load \ k       &k\in 1..h_i,\
    \vc{E}_{i+1} = \vc{E}_{i} \cdot \vc{E}_{i}[k] \\

    \Build \ \s{c} \ n
    &\s{c}:\vc{t}\arrow t,\
    \vc{E}_{i}=E\cdot \vc{e},\ 
    |\vc{e}|=n,\
    \vc{e}:\vc{t},\
    \vc{E}_{i+1} = E \cdot \s{c}(\vc{e}) \\

    \Call \ g \ n
    &g:\vc{t}\arrow t,\
    \vc{E}_{i}=E\cdot \vc{e},\
    |\vc{e}|=n,\
    \vc{e}:\vc{t},\
    \vc{E}_{i+1} = E \cdot g(\vc{e}) \\

    \Tcall \ g \ n
    &g:\vc{t} \arrow \beta,\
    \vc{E}_{i}=E\cdot \vc{e},\
    |\vc{e}|=n,\
    \vc{e}:\vc{t} \\

    \Return         &f:\vc{t}\arrow t,\
    \vc{E}_{i}=E\cdot e,\
    e:t\\

    \Read \ \s{r}   &\s{r}:\w{Ref}(t),\
    \vc{E}_{i+1} = \vc{E}_{i} \cdot x_{f,i}^{t}\\

    \Read \ k       &k\in 1..h_i,\ 
    \vc{E}_{i}[k]:\w{Ref}(t),\
    \vc{E}_{i+1} = \vc{E}_{i} \cdot x_{f,i}^{t} \\ 

    \Write \ \s{r}  &\s{r}:\w{Ref}(t),\
    \vc{E}_{i} = E\cdot e,\ 
    e:t,\
    \vc{E}_{i+1} = E \\

    \Write \ k     &k\in 1..h_i, \
    \vc{E}_{i}[k]:\w{Ref}(t),\ 
    \vc{E}_{i} = E\cdot e,\ 
    e:t,\
    \vc{E}_{i+1} = E \\

    \Yield          &\vc{E}_{i+1}=\vc{E}_{i} \\

    \Next           &\vc{E}_{i+1}=\vc{E}_{i} \\

    \Wait \ j       &\vc{E}_{i}= \vc{E}_{j}\cdot x_{f,j}^{t},\
    \vc{E}_{i+1}=\vc{E}_{j},\
    \vc{\sigma}_i=\vc{\sigma}_j
  \end{array}
  \]}
\end{table}

\begin{example}\label{ex:shape-verification}
  We give the shape of the values on the stack (a side result of the
  shape analysis) for the bytecode obtained from the compilation of
  the function \w{f} defined in Example~\ref{max-value-ex}: 
  \[{\small
    \begin{array}{@{\qquad}l@{\quad}|@{\quad}l@{\qquad}|
        |@{\qquad}l@{\quad}|@{\quad}l@{\qquad}}
      \w{Instruction}       &\w{Shape}
      &\w{Instruction}       &\w{Shape} \\ 
      \hline
      1:\ \Yield               &x 
      &4:\ \Call \ \w{maxl}\ 2  &x \cdot l \cdot x \\
      
      2:\ \Read \ \s{i}        &x 
      &5:\ \Call\ f_{1}\ 1      &x \cdot \w{maxl}(l,x) \\
      
      3:\ \Load \ 1            &x \cdot l 
      &6:\ \Return              &x \cdot \w{f}_{1}(\w{maxl}(l,x))
    \end{array}}
    \]
  
  \noindent Note that the code has no $\Branch$ instruction, hence the
  substitution $\sigma$ is always the identity. Once the shapes are
  generated it is rather straightforward to determine a set of
  constraints that entails the termination of the code and a bound on
  the size of the computed values. For instance, assuming the
  reduction order is a simplification order, it is enough to require
  that $\hatt{f}(x,l)>\w{f}_{1}(\w{maxl}(l,x))$, i.e. the shape of the
  returned value, $\w{f}_{1}(\w{maxl}(l,x))$, is less than the shape
  of the call, $\hatt{f}(x,l)$.
\end{example}

If one can find a reduction order and an assignment satisfying the
constraints generated from the shape analysis then one can show the
termination of the instant and provide bounds on the size of the
computed values. We refrain from developing this part which is
essentially an adaptation of Section~\ref{res-contr} at bytecode
level. Moreover, a detailed treatment of the functional fragment is
available in~\cite{ACDJ04}.  Instead, we state that the shape analysis
is always successful on the bytecode generated by the compilation
function described in Table \ref{compilation} (see
Appendix~\ref{sec:compiled-code-well}).  This should suggest that the
control flow analysis is not overly constraining though it can
certainly be enriched in order to take into account some code
optimisations.

\begin{theorem}\label{compil-success}
  The shape analysis succeeds on the compilation of a well-formed
  program.
\end{theorem}

\section{Conclusion}
The execution of a thread in a cooperative synchronous model can be
regarded as a sequence of instants. One can make each instant simple
enough so that it can be described as a function --- our experiments
with writing sample programs show that the restrictions we impose do
not hinder the expressivity of the language.  Then well-known static
analyses used to bound the resources needed for the execution of
first-order functional programs can be extended to handle systems of
synchronous cooperative threads.  We believe this provides some
evidence for the relevance of these techniques in concurrent/embedded
programming.  We also expect that our approach can be extended to a
richer programming model including more complicated control
structures.\\
The static analyses we have considered do not try to analyse the whole
system. On the contrary, they focus on each thread separately and can
be carried out incrementally.  Moreover, it is quite possible to
perform them at bytecode level.  These characteristics are
particularly interesting in the framework of `mobile code' where
threads can enter or leave the system at the end of each instant as
described in~\cite{Boudol04}.

\paragraph*{Acknowledgements and Publication History}
We would like to thank the referees for their valuable comments.
Thanks to G.~Boudol and F.~Dabrowski for comments and discussions on a
preliminary version of this article that was presented at the 2004
International Conference on Concurrency Theory. In the present paper,
we consider a more general model which includes references as first
class values and requires a reformulation of the control flow
analysis. Moreover, we present a new virtual machine, a number of
examples, and complete proofs not available in the
conference paper.

\appendix

\section{Readers-Writers and Other Synchronisation Patterns}\label{examples}
A simple, maybe the simplest, example of synchronisation and resource
protection is the single place buffer. The buffer (initially empty) is
implemented by a thread listening to two signals. The first on the
register $\s{put}$ to fill the buffer with a value if it is empty, the
second on the register $\s{get}$ to emit the value stored in the
buffer by writing it in the special register $\s{result}$ and flush
the buffer. In this encoding, the register $\s{put}$ is a one place
channel and $\s{get}$ is a signal as in Example~\ref{ex-ch-sig}.
Moreover, owing to the read once condition, we are not able to react
to several put/get requests during the same instant --- only if the
buffer is full can we process one get and one put request in the same
instant.  Note that the value of the buffer is stored on the function
call to $\w{full}(v)$, hence we use function parameters as a kind of
private memory (to compare with registers that model shared memory).

\noindent
\begin{center}
  ${\small \begin{array}[c]{lcl}
    \w{empty}() &=& \w{read}\, \s{put} \,\w{with}\,
    \s{full}(x) \Arrow \w{next} . \w{full}(x)\
    \mid\ [\_] \Arrow \w{empty}() \\

    \w{full}(x) &=&
    \w{read}\, \s{get} \,\w{with}\,
    \s{prst} \Arrow \s{result} := x . \w{yield} .
    \w{empty}()\ 
    \mid\ [\_] \Arrow \w{full}(x)
  \end{array}}$
\end{center}

Another common example of synchronisation pattern is a situation where
we need to protect a resource that may be accessed both by `readers'
(which access the resource without modifying it) and `writers' (which
can access and modify the resource). This form of access control is
common in databases and can be implemented using traditional
synchronisation mechanisms such as semaphores, but this implementation
is far from trivial~\cite{Ode00}.\\
In our encoding, a control thread secures the access to the protected
resource. The other threads, which may be distinguished by their
identity \emph{id} (a natural number), may initiate a request to
access / release the resource by sending a special value on the
dedicated register $\s{req}$. The thread regulating the resource may
acknowledge at most one request per instant and allows the sender of a
request to proceed by writing its $\w{id}$ on the register $\s{allow}$
at the next instant. The synchronisation constraints are as follows:
there can be multiple concurrent readers, there can be only one writer
at any one time, pending write requests have priority over pending
read requests (but do not preempt ongoing read operations).\\
We define a new algebraic datatype for assigning requests:
\[ 
\w{request} = \s{startRead}(\w{nat}) \mid \s{startWrite}(\w{nat}) \mid
\s{endRead} \mid \s{endWrite} \mid \s{none}
\]

The value $\s{startRead}(\w{id})$ indicates a read request from the
thread $\w{id}$, the other constructors correspond to requests for
starting to write, ending to read or ending to write --- the value
\s{none} stands for no requests.  A \s{startRead} operation requires
that there are no pending writes to proceed. In that case we increment
the number of ongoing readers and allow the caller to proceed.  By
contrast, a \s{startWrite} puts the monitor thread in a state waiting
to process the pending write request (function $\w{pwrite}$), which
waits for the number of readers to be null and then allows the thread
that made the pending write request to proceed. An \s{endRead} and
\s{endWrite} request is always immediately acknowledged.\\
The thread protecting the resource starts with the behaviour
$\w{onlyreader}(\s{z})$, defined in Table~\ref{table:onlyreader},
meaning the system has no pending requests for reading or writing. The
behaviour $\w{onlyreader}(x)$ encodes the state of the controller when
there is no pending write and $x$ readers. In a state with $x$ pending
readers, when a \s{startWrite} request from the thread $\w{id}$ is
received, the controller thread switches to the behaviour
$\w{pwrite}(id, x)$, meaning that the thread $\w{id}$ is waiting to
write and that we should wait for $x$ \s{endRead} requests before
acknowledging the request to write.\\
A thread willing to read on the protected resource should repeatedly
try to send its request on the register $\s{req}$ then poll the
register $\s{allow}$, e.g., with the behaviour $\w{askRead}(\w{id}) .
\w{read}\, \s{allow}$ $\w{with}\, \w{id} \Arrow \cdots$ where
$\w{askRead}(\w{id})$ is a shorthand for $\w{read}\, \s{req}
\,\w{with}\, \s{none} \Arrow \s{req} := \s{startRead}(\w{id})$. The
code for a thread willing to end a read session is similar. It is
simple to change our encoding so that multiple requests are stored in
a fifo queue instead of a one place
buffer.

\begin{table}
  \caption{Code for the Readers-Writers pattern}\label{table:onlyreader}
  {\small
    \[
    \begin{array}[t]{lcl}
      \w{onlyreader}(x) & = & 
      
      \w{match}\ x \,\w{with}\, \s{s}(x') \,\w{then}\,
      \w{read}\, \s{req} \,\w{with}\\
      &&\ 
      \begin{array}[t]{l@{}l}
        & \s{endRead} \Arrow  \w{next} . \w{onlyreader}(x')\\
        \mid\ & \s{startWrite}(y) \Arrow \w{next}
        . \w{pwrite}(y, \s{s}(x'))\\
        \mid\ & \s{startRead}(y) \Arrow \w{next} . \s{allow} := y
        . \w{onlyreader}(\s{s}(\s{s}(x')))\\
        \mid\ & [\_] \Arrow \w{onlyreader}(\s{s}(x'))\\
      \end{array}\\
      && \w{else}\, \w{read}\, \s{req} \,\w{with}\\
      &&\ 
      \begin{array}[t]{l@{}l}
        & \s{startWrite}(y) \Arrow \w{next} . \s{allow} := y
        . \w{pwrite}(y, \s{z})\\
        \mid\ & \s{startRead}(y) \Arrow \w{next} . \s{allow} := y
        . \w{onlyreader}(\s{s}(\s{z}))\\
        \mid\ & [\_] \Arrow \w{onlyreader}(\s{z})\\[0.5em]
      \end{array}\\[0.5em]
      
      \w{pwrite}(\w{id}, x) & = & \w{match}\ x \,\w{with}\, \s{s}(x')
      \,\w{then}\\
      &&\quad \w{match}\ x' \,\w{with}\, \s{s}(x'')\,\w{then}\,
      \w{read}\, \s{req} \,\w{with}\\
      &&\quad \
      \begin{array}[t]{l@{}l}
        & \s{endRead} \Arrow \w{next} . \w{pwrite}(\w{id}, \s{s}(x''))\\
        \mid\ & [\_] \Arrow \w{pwrite}(\w{id}, \s{s}(\s{s}(x'')))\\
      \end{array}\\
      &&\quad \w{else}\, \w{read}\, \s{req} \,\w{with}\\
      &&\quad\
      \begin{array}[t]{l@{}l}
        & \s{endRead} \Arrow \w{next} . \s{allow} := \w{id} . 
        \w{pwrite}(\w{id}, \s{z})\\
        \mid\ & [\_] \Arrow \w{pwrite}(\w{id}, \s{s}(\s{z}))\\
      \end{array}\\
      &&\w{else}\, \w{read}\, \s{req} \,\w{with}\\
      &&\
      \begin{array}[t]{l@{}l}
        & \s{endWrite} \Arrow \w{next} . \w{onlyreader}(\s{z})\\
        \mid\ & [\_] \Arrow \w{pwrite}(\w{id}, \s{z})
      \end{array}
    \end{array}
    \]}
\end{table}

\section{Proofs}\label{proofs}
\subsection{Preservation of Control Points Instances}\label{prop-cp-proof}
\begin{proposition}{\ref{prop-cp}}
  Suppose $(B,s,i) \arrow (B',s',i')$ and that for all thread indexes
  $j\in\Z_n$, $B_1(j)$ is an instance of a control point. Then for all
  $j\in\Z_n$, we have that $B'_1(j)$ is an instance of a control
  point.
\end{proposition}

\begin{table}
  \caption{Expression body evaluation 
    and behaviour reduction revised}\label{beh-red-rev}
  {\small
    \[
    \begin{array}{@{\quad}c@{\quad}}
      \\
      
      (\s{e}_0)~\infer{}{(f(\vc{p}),x,\sigma) \eval \sigma(x)} \qquad
      
      (\s{e}_{1})~\infer{}{(f(\vc{p}),\s{r},\sigma)\eval \s{r}} \\

      (\s{e}_{2})~
      \infer{ ( f(\vc{p}), e_i, \sigma) \eval v_i~~i\in 1..n}
      {(f(\vc{p}),\s{c}(\vc{e}),\sigma) \eval \s{c}(\vc{v})} 
      
      ~~

      (\s{e}_3)~\infer{\begin{array}{c}
          ( f(\vc{p}), e_i, \sigma) \eval v_i~~i\in 1..n, \\
          \quad g(\vc{x})=\w{eb},\quad
          (g(\vc{x}),\w{eb},[\vc{v}/\vc{x}]) \eval v
        \end{array}}
      {(f(\vc{p}),g(\vc{e}),\sigma)\eval v} \\

      (\s{e}_4)~\infer{\begin{array}{c}
          \sigma(x)=\s{c}(\vc{v}),\\
          (f([\s{c}(\vc{x})/x]\vc{p}), 
          \w{eb}_{1}, 
          [\vc{v}/\vc{x}]\comp \sigma) \eval v
        \end{array}}
      {\left( f(\vc{p}),
          \begin{array}{c}
            \w{match}~x\\ \w{with}~\s{c}(\vc{x}) \\
            \w{then}~\w{eb}_{1}~\w{else}~\w{eb}_{2}           
          \end{array},
          \sigma \right) \eval v}
      ~~

      (\s{e}_5)~\infer{\begin{array}{c}
          \sigma(x)=\s{d}(\ldots), \\
          (f(\vc{p}), 
          \w{eb}_{2}, 
          \sigma) \eval v
        \end{array}}
      {\left( f(\vc{p}),
          \begin{array}{c}
            \w{match}~x\\\w{with}~\s{c}(\vc{x}) \\
            \w{then}~\w{eb}_{1}~\w{else}~\w{eb}_{2}           
          \end{array},
          \sigma \right) \eval v}\\
      
      (\s{b}_1)~~\infer{}
      { (\hatt{f}(\vc{p}),\w{stop},\sigma, s) \act{S}
        (\hatt{f}(\vc{p}),\w{stop},\sigma, s) }\\
      
      (\s{b}_2)~~\infer{}
      { (\hatt{f}(\vc{p}),\w{yield}.b,\sigma, s) \act{R} (\hatt{f}(\vc{p}),b,\sigma, s)} \\
      
      (\s{b}_3)~~\infer{}
      { (\hatt{f}(\vc{p}),\w{next}.g(\vc{e}),\sigma, s) 
        \act{N} 
        (\hatt{f}(\vc{p}),g(\vc{e}),\sigma, s) }\\
  
      (\s{b}_4)~\infer{\sigma(x)=\s{c}(\vc{v}),\quad
        (\hatt{f}([\s{c}(\vc{x})/x]\vc{p}),
        b_{1}, [\vc{v}/\vc{x}]\comp \sigma, s)
        \act{X} 
        (\hatt{f}_{1}(\vc{p'}), b',\sigma',s')}
      {\left(
          \hatt{f}(\vc{p}),
          \begin{array}{l}
            \w{match}~x~\w{with}~\s{c}(\vc{x}) \\
            \w{then}~b_{1}~\w{else}~b_{2}
          \end{array},
          \sigma, s
        \right) \act{X} 
        (\hatt{f}_{1}(\vc{p'}), b',\sigma',s')}\\

      (\s{b}_5)~\infer{\sigma(x)=\s{d}(\ldots),\quad \s{c}\neq \s{d},\quad 
        (\hatt{f}(\vc{p}),
        b_{2}, \sigma, s)
        \act{X} 
        (\hatt{f}_{1}(\vc{p'}), b',\sigma',s')}
      {\left(
          \hatt{f}(\vc{p}),
          \begin{array}{l}
            \w{match}~x~\w{with}~\s{c}(\vc{x}) \\
            \w{then}~b_{1}~\w{else}~b_{2}
          \end{array},
          s,
          \sigma 
        \right) \act{X} 
        (\hatt{f}_{1}(\vc{p'}), b',\sigma',s')}\\
      
      (\s{b}_6)~~\infer{\mbox{no pattern matches }\w{s}(\sigma(\varrho))}
      {(\hatt{f}(\vc{p}),\w{read}~\varrho~\w{with} \ldots,\sigma, s ) \act{W}
        (\hatt{f}(\vc{p}),\w{read}~\varrho~\w{with} \ldots,\sigma, s ) }\\
      
      (\s{b}_7)~~
      \infer{\sigma_1(p) = s(\sigma(\varrho)),~~ 
        (\hatt{f}([p/y]\vc{p}), b,\sigma_1\comp\sigma ,s) \act{X} 
        (\hatt{f}_{1}(\vc{p'}), b',\sigma',s')}
      {(\hatt{f}(\vc{p}),\w{read}_{\langle y\rangle}\, \varrho \,\w{with}\,
        \dots \mid p \Arrow b
        \mid \dots,\sigma,s) \act{X} 
        (\hatt{f}_{1}(\vc{p'}), b',\sigma',s')}\\
  
      (\s{b}_8)~~
      \infer{\begin{array}{c}
          \sigma \vc{e}\eval \vc{v},
          ~~g(\vc{x})= b, \\
          (\hatt{g}(\vc{x},\vc{y}_g), b, [\vc{v}/\vc{x}], s)\act{X} 
          (\hatt{f}_{1}(\vc{p'}), b',\sigma',s')
        \end{array}}
      {(\hatt{f}(\vc{p}),g(\vc{e}),\sigma,s) \act{X} 
        (\hatt{f}_{1}(\vc{p'}), b',\sigma',s')}\\
      
      (\s{b}_9)~~
     \infer{\sigma e\eval v,~~(\hatt{f}(\vc{p}),b,\sigma,s[v/\sigma(\varrho)]) 
        \act{X} (\hatt{f}_{1}(\vc{p'}), b',\sigma',s')}
     {(\hatt{f}(\vc{p}),\varrho := e.b,\sigma,s) \act{X} 
        (\hatt{f}_{1}(\vc{p'}), b',\sigma',s')}\\
      
      \hbox to \textwidth{}\\[1em]
    \end{array} 
    \]}
\end{table}
  
\Proof
  Let $(f(\vc{p}),\w{be},i)$ be a control point of an expression body
  or of a behaviour. In Table~\ref{beh-red-rev}, we reformulate the
  evaluation and the reduction by replacing expression bodies or
  behaviours by triples $(f(\vc{p}),\w{be},\sigma)$ where
  $(f(\vc{p}),\w{be},i)$ is a control point and $\sigma$ is a
  substitution mapping the variables in $\vc{p}$ to values. By
  convention, we take $\sigma(\s{r}) = \s{r}$ if $\s{r}$ is a register.\\
  We claim that the evaluation and reduction in
  Table~\ref{beh-red-rev} are equivalent to those presented in
  Section~\ref{sct} in the following sense:
  \begin{enumerate}
    
  \item $(f(\vc{p}), e_0,\sigma)\eval v$ iff $\sigma e_0\eval v$.
    
  \item $(\hatt{f}(\vc{p}),b_0,s,\sigma)\act{X}
    (\hatt{g}(\vc{q}),b'_{0},s',\sigma')$ iff $\sigma b_0 \act{X}
    \sigma' b'_{0}$.
  \end{enumerate}
  
  In the following proofs we will refer to the rules in
  Table~\ref{beh-red-rev}. The revised formulation makes clear that if
  $b$ is an instance of a control point and $(b,s) \smash[t]{\act{X}}
  (b',s')$ then $b'$ is an instance.  It remains to check that being
  an instance is a property preserved at the level of system
  reduction. We proceed by case analysis on the last reduction rule
  used in the
  derivation of $(B,s,i) \arrow (B',s',i')$.\\
  \Proofitem{$(\s{s}_1)$} One of the threads performs one step.  The
  property follows by the analysis on behaviours.\\
  \Proofitem{$(\s{s}_2)$} One of the threads performs one step.
  Moreover, the threads in waiting status take the $[\_] \Arrow
  g(\vc{e})$ branch of the \w{read} instructions that were blocking.
  A thread $\w{read}~\varrho~\ldots \mid [\_]\Arrow g(\vc{e})$ in
  waiting status is an instance of a control point $(\hatt{f}(\vc{p}),
  \w{read}~\varrho~\ldots \mid [\_]\Arrow g(\vc{e}_{0}),j)$. By
  $(\cl{C}_7)$, $(\hatt{f}(\vc{p}), g(\vc{e}_{0}),2)$ is a control
  point, and
  $g(\vc{e})$ is one of its instances.\qed

\subsection{Evaluation of Closed Expressions}\label{exp-eval-proof}
\begin{proposition}{\ref{exp-eval}}
  Let $e$ be a closed expression. Then there is a value $v$ such that
  $e \eval v$ and $e \geq v$ with respect to the reduction order.
\end{proposition}

As announced, we refer to the rules in Table~\ref{beh-red-rev}.  We
recall that the order $>$ or $\geq$ refers to the reduction order that
satisfies the constraints of index $0$.  We start by proving the
following working lemma.

\begin{lemma}\label{exp-lemma}
  For all well formed triples, $(f(\vc{p}),\w{eb},\sigma)$, there is a
  value $v$ such that $(f(\vc{p}),\w{eb},\sigma)\eval v$.  Moreover,
  if $\w{eb}$ is an expression then $\sigma(\w{eb})\geq v$ else
  $f(\sigma\vc{p}) \geq v$.
\end{lemma}
\Proof
  We proceed by induction on the pair $(f(\sigma\vc{p}),\w{eb})$
  ordered lexicographically from left to right. The first argument is
  ordered according to the reduction order and the second according to
  the structure of the expression body.\\
  \Proofitemm{\w{eb}\equiv x.} We apply rule $(\s{e}_0)$ and
  $\sigma(x)\geq \sigma(x)$.\\
  \Proofitemm{\w{eb}\equiv \s{r}.} We apply rule $(\s{e}_1)$ and
  $\sigma(\s{r})=\s{r}\geq \s{r}$.\\
  \Proofitemm{\w{eb}\equiv \s{c}(e_1,\ldots,e_n).} We apply rule
  $(\s{e}_2)$.  By inductive hypothesis, $(f(\vc{p}),e_i,\sigma)\eval
  v_i$ for $i\in 1..n$ and $\sigma e_i\geq v_i$. By definition of
  reduction order, we derive $\sigma(\s{c}(e_1,\ldots,e_n))\geq
  \s{c}(v_1,\ldots,v_n)$.\\
  \Proofitemm{\w{eb}\equiv f(e_1,\ldots,e_n).} We apply rule
  $(\s{e}_3)$.  By inductive hypothesis, $(f(\vc{p}),e_i,\sigma)\eval
  v_i$ for $i\in 1..n$ and $\sigma e_i\geq v_i$.  By the definition of
  the generated constraints $f(\vc{p})>g(\vc{e})$, which by definition
  of reduction order implies that $f(\sigma\vc{p})>g(\sigma\vc{e})
  \geq g(\vc{v})= g([\vc{v}/\vc{x}]\vc{x})$.  Thus by inductive
  hypothesis, $g(\vc{x},\w{eb},[\vc{v}/x])\eval v$.  We conclude by
  showing by case analysis that $g(\sigma \vc{e})\geq v$.
  \begin{itemize}
  \item $\w{eb}$ is an expression. By the constraint we have
    $g(\vc{x})> \w{eb}$, and by inductive hypothesis
    $[\vc{v}/\vc{x}]\w{eb}\geq v$.  So $g(\sigma\vc{e}) \geq g(\vc{v})
    > [\vc{v}/\vc{x}]\w{eb}\geq v$.
  
  \item $\w{eb}$ is not an expression. Then by inductive hypothesis,
    $g(\vc{v}) \geq v$ and we know $g(\sigma\vc{e})\geq g(\vc{v})$.
  \end{itemize}
  \Proofitemm{\w{eb}\equiv \w{match}~x~\w{with}~\s{c}(\vc{x})\ldots\,.}
  We distinguish two cases.
  \begin{itemize}
  \item $\sigma(x)=\s{c}(\vc{v})$.  Then rule $(\s{e}_4)$ applies.
    Let $\sigma'=[\vc{v}/\vc{x}]\comp \sigma$.  Note that $\sigma'
    ([\s{c}(\vc{x})/x]\vc{p}) = \sigma \vc{p}$.  By inductive
    hypothesis, we have that $(f([\s{c}(\vc{x})/x]\vc{p}),\w{eb}_1,
    \sigma')\eval v$.  We show by case analysis that
    $f(\sigma\vc{p})\geq v$.
    
    \begin{itemize}
    \item $\w{eb}_1$ is an expression.  By inductive hypothesis,
      $\sigma' (\w{eb}_1) \geq v$.  By the constraint,
      $f([\s{c}(\vc{x})/x]\vc{p}) > \w{eb}_1$.  Hence, $f(\sigma
      \vc{p}) = f(\sigma'[\s{c}(\vc{x})/x]\vc{p}) >
      \sigma'(\w{eb}_1)$.
    
    \item $\w{eb}_2$ is not an expression.  By inductive hypothesis,
      we have that $f(\sigma \vc{p})$ equals
      $f(\sigma'[\s{c}(\vc{x})/x]\vc{p}) \geq v$.
    \end{itemize}
  
  \item $\sigma(x)=\s{d}(\ldots)$ with $\s{c} \neq \s{d}$. Then rule
    $(\s{e}_5)$ applies and an argument simpler than the one above
    allows to conclude. \qed
  \end{itemize}

Relying on Lemma~\ref{exp-lemma} we can now prove
Proposition~\ref{exp-eval}, that if $e$ is a closed expression and
$e\eval v$ then $e\geq v$ in the reduction order.
\Proof
  We proceed by induction on the structure of $e$.\\
  \Proofitem{$e$ is value $v$.}  Then $v\eval v$ and $v\geq v$.\\
  \Proofitem{$e\equiv \s{c}(e_1,\ldots,e_n)$.}  By inductive
  hypothesis, $e_i\eval v_i$ and $e_i\geq v_i$ for $i\in 1..n$.  By
  definition of reduction order, $\s{c}(\vc{e})\geq \s{c}(\vc{v})$.\\
  \Proofitem{$e\equiv f(e_1,\ldots,e_n)$.}  By inductive hypothesis,
  $e_i\eval v_i$ and $e_i\geq v_i$ for $i\in 1..n$.  Suppose
  $f(\vc{x})=\w{eb}$. By Lemma~\ref{exp-lemma},
  $(f(\vc{x}),\w{eb},[\vc{v}/\vc{x}])\eval v$ and either
  $f(\vc{v})\geq v$ or $f(\vc{x})>\w{eb}$ and $\sigma(\w{eb})\geq v$.
  We conclude by a simple case analysis. \qed

\subsection{Progress}\label{beh-red-proof}
\begin{proposition}{\ref{beh-red}}
  Let $b$ be an instance of a control point. Then for all stores $s$,
  there exists a store $s'$ and a status $X$ such that
  $(b,s)\act{X}(b',s')$.
\end{proposition}

\Proof
  We start by defining a suitable well-founded order.  If $b$ is a
  behaviour, then let $\w{nr}(b)$ be the maximum number of reads that
  $b$ may perform in an instant.  Moreover, let $\w{ln}(b)$ be the
  {\em length} of $b$ inductively defined as follows:
  \[{\small
  \begin{array}[c]{l}
    \begin{array}{c}
      \w{ln}(\Stop)=\w{ln}(f(\vc{e}))=0
      \quad
      \w{ln}(\w{yield}.b)= \w{ln}(\varrho := e.b)= 1+\w{ln}(b)
      \quad
      \w{ln}(\w{next}.f(\vc{e}))=2  \\

      \w{ln}(\w{match}~x~\w{with}~\s{c}(\vc{x})~\w{then}~b_1~\w{else}~b_2) = 
      1+\w{max}(\w{ln}(b_1),\w{ln}(b_2)) \\
      
      \w{ln}(\w{read}~\varrho~\w{with} \ldots \mid p_i \Arrow b_i 
      \mid \ldots \mid [\_] \Arrow f(\vc{e})) = 1 + \w{max}(\ldots, \w{ln}(b_i),\ldots)
    \end{array}
  \end{array}}
  \]
  If the behaviour $b$ is an instance of the control point
  $\gamma\equiv (\hatt{f}(\vc{p}),b_0,i)$ via a substitution $\sigma$
  then we associate with the pair $(b,\gamma)$ a measure:
  \[
  \mu(b,\gamma) \ \eqdef\ 
  (\w{nr}(b),\hatt{f}(\sigma\vc{p}),\w{ln}(b))~.
  \]
  
  We assume that measures are lexicographically ordered from left to
  right, where the order on the first and third component is the
  standard order on natural numbers and the order on the second
  component is the reduction order considered in study of the
  termination conditions. This is a well-founded order. Now we show
  the assertion by induction on $\mu(b,\gamma)$.  We proceed by case
  analysis on the structure of $b$.\\
  \Proofitem{$b \equiv \w{stop}$.} Rule $(\s{b}_1)$ applies, with
  $X=S$, and the measure stays constant.\\
  \Proofitem{$b \equiv \w{yield}.b'$.}~Rule $(\s{b}_2)$ applies, with
  $X=R$, and the measure decreases because $\w{ln}(b)$ decreases.\\
  \Proofitem{$b \equiv \w{next}.b'$.}~Rule $(\s{b}_3)$ applies, with
  $X=N$, and the measure decreases because $\w{ln}(b)$ decreases.\\
  \Proofitem{$b \equiv \w{match}\, \ldots\,$.} Rules $(\s{b}_4)$ or
  $(\s{b}_5)$ apply and the measure decreases because $\w{ln}(b)$
  decreases.\\
  \Proofitem{$b \equiv \w{read}\, \dots$.} If no pattern matches then
  rule $(\s{b}_6)$ applies and the measure is left unchanged. If a
  pattern matches then rule $(\s{b}_7)$ applies and the measure
  decreases because $\w{nr}(b)$ decreases and then the induction
  hypothesis applies.\\
  \Proofitem{$b \equiv g(\vc{e})$.} Rule $(\s{b}_8)$ applies to
  $(\hatt{f}(\vc{p}),g(\vc{e}_0),\sigma)$, assuming
  $\vc{e}=\sigma\vc{e}_0$. By Proposition~\ref{exp-eval}, we know that
  $\vc{e}\eval \vc{v}$ and $\vc{e}\geq \vc{v}$ in the reduction order.
  Suppose $g$ is associated to the declaration $g(\vc{x})= b$. The
  constraint associated with the control point requires
  $\hatt{f}(\vc{p})>\hatt{g}(\vc{e}_0,\vc{y}_g)$. Then using the
  properties of reduction orders we observe:
  \[
  \begin{array}{lllll}
    \hatt{f}(\sigma\vc{p}) 
    &> \hatt{g}(\sigma \vc{e}_0,\vc{y}_g) 
    &= \hatt{g}(\vc{e},\vc{y}_g)   
    &\geq \hatt{g}(\vc{v},\vc{y}_g)  
  \end{array}
  \]
  Thus the measure decreases because
  $\hatt{f}(\sigma\vc{p})>\hatt{g}(\vc{v},\vc{y}_g)$, and then the
  induction hypothesis applies.\\
  \Proofitem{$b \equiv \varrho := e.b'$.} By Proposition~\ref{exp-eval}, we
  have $e\eval v$. Hence rule $(\s{b}_9)$ applies, the measure
  decreases because $\w{ln}(b)$ decreases, and then the induction
  hypothesis applies.\qed

\begin{remark}\label{mu-rmk}
  We point out that in the proof of proposition \ref{beh-red}, if
  $X=R$ then the measure decreases and if $X\in\set{N,S,W}$ then the
  measure decreases or stays the same. We use this observation in the
  following proof of Theorem \ref{sys-ter}.
\end{remark} 

\subsection{Termination of the Instant}\label{sys-ter-proof}
\begin{theorem}{\ref{sys-ter}}
  All sequences of system reductions involving only rule $(\s{s}_1)$
  are finite.
\end{theorem}

\Proof
  We order the status of threads as follows: $R>N,S,W$.  With a
  behaviour $B_1(i)$ coming with a control point $\gamma_i$, we
  associate the pair $\mu'(i)=(\mu(B_1(i),\gamma_i),B_2(i))$ where
  $\mu$ is the measure defined in the proof of Proposition
  \ref{beh-red}. Thus $\mu'(i)$ can be regarded as a quadruple with a
  lexicographic order from left to right. With a system $B$ of $n$
  threads we associate the measure $\mu_B \eqdef
  (\mu'(0),\ldots,\mu'(n-1))$ that is a tuple. We compare such tuples
  using the product order. We prove that every system reduction
  sequence involving only rule ($\s{s}_1$) terminates by proving that
  this measure decreases during reduction. We recall the rule below:
  \[{\small
  \infer{(B_1(i),s) \act{X} (b',s'),~~ B_2(i)=R,~~ B'= B[(b',X)/i],~~
    \cl{N}(B',s',i) = k} {(B,s,i) \arrow (B'[(B'_{1}(k),R)/k],s',k)}}
  \]
  Let $B''=B'[(B'_{1}(k),R)/k]$. We proceed by case analysis on $X$
  and $B'_{2}(k)$.\\
  If $B'_{2}(k) = R$ then $\mu'(k)$ is left unchanged. The only other
  case is $B'_{2}(k) = W$. In this case the conditions on the
  scheduler tell us that $i\neq k$. Indeed, the thread $k$ must be
  blocked on a $\w{read}\ \s{r}$ instruction and it can only be
  scheduled if the value stored in $\s{r}$ has been modified, which
  means than some other thread than $k$ must have modified \s{r}. For
  the same reason, some pattern in the \w{read}\ \s{r} instruction of
  $B_1(k)$ matches $s'(\s{r})$, which means that the number of reads
  that $B_1(k)$ may perform in the current instant decreases and that
  $\mu'(k)$ also decreases.\\
  By hypothesis we have $(B_1(i),s) \act{X} (b',s')$, hence by Remark
  \ref{mu-rmk}, $\mu'(i)$ decreases or stays the same. By the previous
  line of reasoning $\mu'(k)$ decreases and the other measures
  $\mu'(j)$ stay the same. Hence the measure $\mu_B$ decreases, as
  needed.\qed

\subsection{Bounding the Size of Values for Threads}\label{thread-bound-proof}
\begin{theorem}{\ref{thread-bound}}
  Given a system of synchronous threads $B$, suppose that at the
  beginning of the instant $B_{1}(i)=f(\vc{v})$ for some thread index
  $i$. Then the size of the values computed by the thread $i$ during
  an instant is bounded by $q_{\hatt{f}(\vc{v},\vc{u})}$ where
  $\vc{u}$ are the values contained in the registers at the time they
  are read by the thread (or some constant value, if they are not read
  at all).
\end{theorem}
In Table~\ref{beh-red-rev}, we have defined the reduction of
behaviours as a {\em big step} semantics. In Table~\ref{small-step} we
reformulate the operational semantics following a {\em small step}
approach. First, note that there are no rules corresponding to
$(\s{b}_1)$, $(\s{b}_3)$ or $(\s{b}_6)$ since these rules either
terminate or suspend the computation of the thread in the instant.
Second, the reduction makes abstraction of the memory and the
scheduler. Instead, the reduction relation is parameterized on an
assignment $\delta$ associating values with the labels of the read
instructions. \\
The assignment $\delta$ is a kind of {\em oracle} that
provides the thread with the finitely many values (because of the read
once condition) it may read within the current instant. The assignment 
$\delta$ provides a {\em safe} abstraction of the store $s$ 
used in the transition rules of Table~\ref{beh-red-rev}. 
Note that the resulting system represents {\em more} reductions than
can actually occur in the original semantics within an instant.
Namely, a thread can write a value $v$ in $\s{r}$ and then proceed to
read from $\s{r}$ a value different from $v$ without yielding the
control. This kind of reduction is impossible in the original
semantics. However, since we do not rely on a precise monitoring of
the values written in the store, this loss of precision does not
affect our analysis.\\
Next we prove that if $(\hatt{f}(\vc{p}),b,\sigma) \arrow_{\delta}
(\hatt{g}(\vc{q}),b',\sigma')$ then $q_{\hatt{f}(\sigma'' \comp
  \sigma(\vc{p}))}\geq$ $q_{\hatt{g}(\sigma'(\vc{q}))}$ over the
non-negative reals, where $\sigma''$ is either the identity or the
restriction of $\delta$ to the label of the \w{read} instruction in
case $(\s{b'}_{7})$.

\begin{table}
  \caption{Small step reduction within an instant}\label{small-step}
  {\small
    \[
    \begin{array}{ll}

      (\s{b'}_2)
      &(\hatt{f}(\vc{p}),\w{yield}.b,\sigma) \arrow_{\delta}
      (\hatt{f}(\vc{p}),b,\sigma) \\  
    
      (\s{b'}_4)
      &\bigl(
        \hatt{f}(\vc{p}),
        \begin{array}{l@{}}
          \w{match}~x~\w{with}~\s{c}(\vc{x}) \\
          \w{then}~b_{1}~\w{else}~b_{2}
        \end{array},
        \sigma \bigr)
      \arrow_{\delta}
      (\hatt{f}([\s{c}(\vc{x})/x]\vc{p}),
      b_{1}, [\vc{v}/\vc{x}]\comp \sigma) ~~\mbox{ if }(1) \\
      
      (\s{b'}_5) &\bigl(
        \hatt{f}(\vc{p}),
        \begin{array}{l@{}}
          \w{match}~x~\w{with}~\s{c}(\vc{x}) \\
          \w{then}~b_{1}~\w{else}~b_{2}
        \end{array},
        \sigma\bigr)
      \arrow_{\delta}
      (\hatt{f}(\vc{p}),
      b_{2}, \sigma) ~~\mbox{ if }
      \sigma(x)=\s{d}(\ldots),\s{c}\neq \s{d} \\
    
      (\s{b'}_{7})
      &(\hatt{f}(\vc{p}),\w{read}_{\langle y\rangle}\, \varrho \,\w{with}\,
      \dots \mid p\Arrow b \mid \dots,\sigma) 
      \arrow_{\delta}
      (\hatt{f}([p/y]\vc{p}), b,\sigma_1\comp\sigma) 
      ~~\mbox{ if }(2) \\
    
      (\s{b'}_{8})
      &(\hatt{f}(\vc{p}),g(\vc{e}),\sigma) 
      \arrow_{\delta}
      (\hatt{g}(\vc{x},\vc{y}_g), b, [\vc{v}/\vc{x}])
      ~~\mbox{ if }
      \sigma\vc{e}\eval \vc{v} \mbox{ and } g(\vc{x})= b \\
    
      (\s{b'}_{9})
      &(\hatt{f}(\vc{p}),\varrho:=e.b,\sigma) 
      \arrow_{\delta}
      (\hatt{f}(\vc{p}),b,\sigma) 
      ~~\mbox{ if } \sigma e\eval v   \\

      &\mbox{where: }(1)\equiv \sigma(x)=\s{c}(\vc{v})
      \mbox{ and }(2)\equiv \sigma_1(p) = \delta(y).
    \end{array} 
    \]}
\end{table}

\Proof
  By case analysis on the small step rules. Cases
  $(\s{b'}_{2})$, $(\s{b'}_{5})$ and $(\s{b'}_{9})$ are immediate.\\
  \Proofitemm{(\s{b'}_{4})} The assertion follows by a straightforward
  computation on
  substitutions.\\
  \Proofitemm{(\s{b'}_{7})} Then
  $\sigma''(y)=\delta(y)=[\sigma_{1}(p)/y]$ and recalling that
  patterns are linear, we note that: $\hatt{f}((\sigma''\comp
  \sigma)(\vc{p})) =
  \hatt{f}((\sigma_{1} \comp \sigma)[p/y](\vc{p}))$.\\
  \Proofitemm{(\s{b'}_{8})} By the properties of
  quasi-interpre\-ta\-tions, we know that $q_{\sigma(\vc{e})}\geq
  q_{\vc{v}}$.  By the constraints generated by the control points, we
  derive that $q_{\hatt{f}(\vc{p})}\geq q_{\hatt{g}(\vc{e},\vc{y}_g)}$
  over the non-negative reals. By the substitutivity property of
  quasi-interpretations, this implies that $q_{\hatt{f}(\sigma
    (\vc{p}))}\geq q_{\hatt{g}(\sigma(\vc{e},\vc{y}_g))}$.  Thus we
  derive, as required: $q_{\hatt{f}(\sigma(\vc{p}))} \geq
  q_{\hatt{g}(\sigma(\vc{e},\vc{y}_g))} \geq
  q_{\hatt{g}(\vc{v},\vc{y}_g)}$. \qed

It remains to support our claim that all values computed by the thread
$i$ during an instant have a size bounded by $q_{f(\vc{v},\vc{u})}$
where $\vc{u}$ are either the values read by the thread or some
constant value.

\Proof
  By inspecting the shape of behaviours we see that a thread {\em
    computes values} either when writing into a register or in
  recursive calls. We consider in turn the two cases.\\
  \Proofitem{\mbox{Writing}} Suppose
  $(\hatt{f}(\vc{p},\vc{y}_f),b,\sigma) \arrow_{\delta}^*
  (\hatt{g}(\vc{q}), \varrho := e.b',\sigma')$ by performing a series
  of reads recorded by the substitution $\sigma''$.  Then the
  invariant we have proved above implies that:
  $q_{\hatt{f}((\sigma''\comp \sigma)(\vc{p},\vc{y}_f))} \geq
  q_{\hatt{g}(\sigma'\vc{q})}$ over the non-negative reals.  If some
  of the variables in $\vc{y}_f$ are not instantiated by the
  substitution $\sigma''$, then we may replace them by some constant.
  Next, we observe that the constraint of index $1$ associated with
  the control point requires that $q_{\hatt{g}(\vc{q})}\geq q_{e}$ and
  that if $\sigma(e) \eval v$ then this implies
  $q_{\hatt{g}(\sigma'(\vc{q}))}\geq q_{\sigma'(e)} \geq q_{v} \geq |v|$.\\
  \Proofitem{\mbox{Recursive call}} Suppose
  $(\hatt{f}(\vc{p},\vc{y}_f),b,\sigma) \arrow_{\delta}^*
  (\hatt{g}(\vc{q}),h(\vc{e}),\sigma')$ by performing a series of
  reads recorded by the substitution $\sigma''$.  Then the invariant
  we have proved above implies that: $q_{\hatt{f}((\sigma'' \comp
    \sigma)(\vc{p},\vc{y}_f))} \geq q_{\hatt{g}(\sigma'(\vc{q}))}$
  over the non-negative reals.  Again, if some of the variables in
  $\vc{y}_f$ are not instantiated by the substitution $\sigma''$, then
  we may replace them by some constant value.  Next we observe that
  the constraint of index $0$ associated with the control point
  requires that $q_{\hatt{g}(\vc{q})}\geq
  q_{\hatt{h}(\vc{e},\vc{y}_h)}$.  Moreover, if $\sigma'(\vc{e}) \eval
  \vc{v}$ then $q_{\hatt{g}(\sigma'(\vc{q}))}\geq q_{\hatt{h}(\sigma'
    (\vc{e},\vc{y}_h))} \geq q_{\hatt{h}(\vc{v},\vc{y}_h)} \geq
  q_{v_{i}} \geq |v_{i}|$, where $v_i$ is any of the values in
  $\vc{v}$. The last inequation relies on the monotonicity property of
  assignments, see property (3) in Definition~\ref{assignment}, that
  is $q_{\hatt{h}}(z_1,\ldots,z_n) \geq z_j$ for all $j \in 1..n$.\qed

\subsection{Bounding the Size of Values for Systems}\label{sys-bound-proof}
\begin{corollary}{\ref{sys-bound}}
  Let $B$ be a system with $m$ distinct read instructions and $n$ threads.  Suppose
  $B_1(i)=f_i(\vc{v}_i)$ for $i\in \Z_n$. Let $c$ be a bound of the
  size of the largest parameter of the functions $f_{i}$ and the
  largest default value of the registers. Suppose $h$ is a function
  bounding all the quasi-interpretations, that is, for all the
  functions $\hatt{f}_{i}$ we have $h(x)\geq
  \smash[b]{q_{\hatt{f}_{i}}(x,\ldots,x)}$ over the non-negative
  reals. Then the size of the values computed by the system $B$ during
  an instant is bounded by $h^{n\cdot m + 1}(c)$.
\end{corollary}

\Proof
  Because of the read once condition, during an instant a system can
  perform a (successful) read at most $n\cdot m$ times.  We proceed by
  induction on the number $k$ of reads the system has performed so far
  to prove that the size of the values is bounded by $h^{k+1}(c)$.\\
  \Proofitem{$k=0$} If no read has been performed, then Theorem
  \ref{thread-bound} can be applied to show that all values have size
  bound by $h(c)$.\\
  \Proofitem{$k>0$} Inductively, the size of the values in the
  parameters and the registers is bounded by $h^{k}(c)$. Theorem
  \ref{thread-bound} says that all the values that can be computed
  before performing a new read have a size bound by $h(h^{k}(c)) =
  h^{k+1}(c)$.\qed

\subsection{Combination of LPO and Polynomial Quasi-interpretations}\label{pspace-bound-proof}
\begin{theorem}{\ref{pspace-bound}}
  If a system $B$ terminates by LPO and admits a polynomial
  quasi-interpre\-ta\-tion then the computation of the system in an
  instant runs in space polynomial in the size of the parameters of
  the threads at the beginning of the instant.
\end{theorem}

\Proof
  We can always choose a polynomial for the function $h$ in corollary
  \ref{sys-bound}. Hence, $h^{nm+1}$ is also a polynomial.  This shows
  that the size of all the values computed by the system is bounded by
  a polynomial. The number of values in a frame depends on the number
  of formal parameters and local variables and it can be statically
  bound.  It remains to bound the number of frames on the stack.  Note
  that behaviours are tail recursive. This means that the stack of
  each thread contains a frame that never returns a value plus
  possibly a sequence of frames that relate to the evaluation of
  expressions.\\
  From this point on, one can follow the proof in~\cite{BMM01}.  The
  idea is to exploit the characteristics of the LPO order: a nested
  sequence of recursive calls $f_1(\vc{v}_{1}),\ldots,
  f_n(\vc{v}_{n})$ must satisfy $f_1(\vc{v}_{1}) > \cdots >
  f_n(\vc{v}_{n})$, where $>$ is the LPO order on terms.  Because of
  the polynomial bound on the size of the values and the characteristics of
  the LPO on constructors, one can provide a
  polynomial bound on the length of such strictly decreasing sequences
  and therefore a polynomial bound on the size of the stack needed to
  execute the system. \qed

\subsection{Compiled Code is Well-shaped}\label{sec:compiled-code-well}
\begin{theorem}{\ref{compil-success}}
  The shape analysis succeeds on the compilation of a well-formed
  program.
\end{theorem}

Let $\w{be}$ be either a behaviour or an expression body, $\eta$ be a
sequence of variables, and $E$ be a sequence of expressions.  We say
that the triple $(\w{be},\eta,E)$ is {\em compatible} if for all
variables $x$ free in $\w{be}$, the index $i(x,\eta)$ is defined and
if $\eta[k]=x$ then $E[k]=x$.  Moreover, we say that the triple is
{\em strongly compatible} if it is compatible and $|\eta|=|E|$.  In
the following we will neglect typing issues that offer no particular
difficulty. First we prove the following lemma.

\begin{lemma}\label{exp-compil}
  If $(e,\eta,E)$ is compatible then the shape analysis of
  $C'(e,\eta)$ starting from the shape $E$ succeeds and produces a
  shape $E\cdot e$. 
\end{lemma}
\Proof
  By induction on the structure of $e$.\\
  \Proofitemm{e\equiv x} Then $C'(x,\eta)= \Load \ i(x,\eta)$.  We
  know that $i(x,\eta)$ is defined and $\eta[k]=x$ implies $E[k]=x$.
  So the shape analysis succeeds and produces $E \cdot x$.\\
  \Proofitemm{e\equiv \s{c}(e_{1},\ldots,e_{n})} Then
  $C'(\s{c}(e_{1},\ldots,e_{n}), \eta) = C'(e_{1},\eta) \cdots
  C'(e_{n},\eta) (\Build \ \s{c} \ n)$.  We note that if $e'$ is a
  subexpression of $e$, $e''$ is another expression, and $(e,\eta,E)$
  is compatible then $(e',\eta,E\cdot e'')$ is compatible too.  Thus
  we can apply the inductive hypothesis to $e_{1},\ldots,e_{n}$ and
  derive that the shape analysis of $C'(e_{1},\eta)$ starting from $E$
  succeeds and produces $E\cdot e_{1}$,\ldots, and the shape analysis
  of $C'(e_{n},\eta)$ starting from $E\cdot e_{1} \cdots e_{n-1}$
  succeeds and produces $E\cdot e_{1}\cdots e_{n}$. Then by the
  definition of shape analysis of $\Build$ we can conclude.\\
  \Proofitemm{e\equiv f(e_{1},\ldots,e_{n})} An argument similar to
  the one above applies.\qed

Next we generalise the lemma to behaviours and expression bodies.

\begin{lemma}\label{shape-compil-beh}
  If $(\w{be},\eta,E)$ is strongly compatible then the shape analysis
  of $C(\w{be},\eta)$ starting from the shape $E$ succeeds.
\end{lemma}

\Proof
  \Proofitemf{\w{be}\equiv e} We have that $C(e,\eta)= C'(e,\eta)\cdot
  \Return$ and the shape analysis on $C'(e,\eta)$ succeeds, producing
  at least one expression.\\
  \Proofitemm{\w{be}\equiv
    \w{match}~x~\w{with}~\s{c}(\vc{y})~\w{then}~\w{eb}_{1}~\w{else}~\w{eb}_2}
  Following the definition of the compilation function, we distinguish
  two cases:
  \begin{itemize}
  \item $\eta\equiv \eta'\cdot x$: Then $C(\w{be},\eta) = (\Branch \ 
    \s{c} \ j)\cdot C(\w{eb}_{1},\eta'\cdot \vc{y})\cdot (j:
    C(\w{eb}_{2},\eta)\ )$.  By the hypothesis of strong
    compatibility, $E\equiv E'\cdot x$ and by definition of shape
    analysis on $\Branch$ we get on the \w{then} branch a shape
    $[\s{c}(\vc{y})/x]E' \cdot \vc{y}$ up to variable renaming.  We
    observe that $(\w{eb}_{1}, \eta'\cdot \vc{y}, [\s{c}(\vc{y})/x]E'
    \cdot \vc{y})$ are strongly compatible (note that here we rely on
    the fact that $\eta'$ and $E'$ have the same length). Hence, by
    inductive hypothesis, the shape analysis on $C(\w{eb}_{1},
    \eta'\cdot \vc{y})$ succeeds.  As for the else branch, we have a
    shape $E'\cdot x$ and since $(\w{eb}_{2}, \eta'\cdot x ,E'\cdot
    x)$ are strongly compatible we derive by inductive hypothesis that
    the shape analysis on $C(\w{eb}_{2},\eta)$ succeeds.
    
  \item $\eta\not\equiv \eta'\cdot x$: The compiled code starts with
    $(\Load \ i(x,\eta))$ which produces a shape $E\cdot x$. Then the
    analysis proceeds as in the previous case.
  \end{itemize}
  
  \Proofitemm{\w{be}\equiv \w{stop}} The shape analysis succeeds.\\
  \Proofitemm{\w{be}\equiv f(e_1,\ldots,e_n)} By lemma
  \ref{exp-compil}, we derive that the shape analysis of
  $C'(e_1,\eta)\cdot$ $\ldots \cdot C'(e_n,\eta)$ succeeds and
  produces $E\cdot e_1\cdots e_n$. We conclude applying the definition
  of the shape
  analysis for $\Tcall$.\\
  \Proofitemm{\w{be}\equiv \w{yield}.b} The instruction $\Yield$ does
  not change the shape and we can apply the inductive hypothesis on
  $b$.\\
  \Proofitemm{\w{be}\equiv \w{next}.g(\vc{e})} The instruction $\Next$
  does not change the shape and we can apply the inductive hypothesis
  on $g(\vc{e})$.\\
  \Proofitemm{\w{be}\equiv \varrho := e.b} By lemma \ref{exp-compil}, we
  have the shape $E\cdot e$.  By definition of the shape analysis on
  $\Write$, we get back to the shape $E$ and then we apply the
  inductive hypothesis on $b$.\\
  \Proofitemm{\w{be}\equiv \w{match}\ldots} The same argument as for
  expression bodies applies.\\
  \Proofitemm{\w{be}\equiv \w{read}\ \varrho \ \w{with}\ 
    \s{c}_{1}(\vc{y}_{1}) \Arrow b_1 \mid \ldots \mid
    \s{c}_{n}(\vc{y}_{n}) \Arrow b_n \mid [\_]\Arrow g(\vc{e})} 
We recall that the compiled code is:
  \[{\small
  \begin{array}{c}
    j_{0}:(\Read \ i(\varrho,\eta)) \cdot
    (\Branch \ \s{c}_{1} \ j_{1}) \cdot C(b_{1},\eta \cdot \vc{y}_{1})\cdots \\
    j_{n-1}:(\Branch \ \s{c}_{n} \ j_{n})\cdot 
    C(b_{n},\eta\cdot \vc{y}_{n})\cdot 
    j_{n}:(\Wait\ j_{0}) \cdot C(g(\vc{e}),\eta) 
  \end{array}}
  \]
    The $\Read$ instruction produces a shape $E\cdot y$.  Then if a
  positive branch is selected, we have a shape $E\cdot \vc{y}_k$ for
  $k\in 1..n$.  We note that the triples $(b_k, \eta \cdot \vc{y}_k, E
  \cdot \vc{y}_k)$ are strongly compatible and therefore the inductive
  hypothesis applies to $C(b_{k},\eta\cdot \vc{y}_{k})$ for $k\in
  1..n$.  On the other hand, if the last default branch $[\_]$ is
  selected then by definition of the shape analysis on $\Wait$ we get
  back to the shape $E$ and again the inductive hypothesis applies to
  $C(g(\vc{e}),\eta)$. The case where a pattern can be a variable is 
  similar.\\
  \Proofitemm{} To conclude the proof we notice that for every
  function definition $f(\vc{x})=\w{be}$, taking $\eta=\vc{x}=\vc{E}$
  we have that $(\w{be},\eta,E)$ are strongly compatible and thus by
  lemma \ref{shape-compil-beh} the shape analysis succeeds on
  $C(\w{be},\eta)$ starting from $E$. \qed

\end{document}